\documentclass{amsart}
\usepackage{longtable}
\usepackage{amsmath}
\usepackage{graphicx}
\usepackage{xcolor}
\usepackage{amssymb}
\usepackage[numbers]{natbib}
\usepackage[utf8]{inputenc}
\usepackage{array} 
\usepackage{booktabs}
\usepackage{geometry}
\usepackage{mathrsfs}
\usepackage{amsmath}
\usepackage{float}

\usepackage{subfigure} 
\usepackage{threeparttable}

\newcommand{\thicktilde}[1]{\mathbf{\tilde{\text{$#1$}}}}
\usepackage{amsmath,array}
\usepackage{physics}
\usepackage{multirow}

\usepackage{adjustbox}

\theoremstyle{definition}

\theoremstyle{remark}

\numberwithin{equation}{section}

\begin{document}

\title[]{ Interval Estimation of Relative Risks for Combined Unilateral and Bilateral Correlated Data}%
\author{Kejia Wang and Chang-Xing Ma}%

\address{Department of Biostatistics, University at Buffalo, New York 14214, USA}%
\email{cxma@buffalo.edu}%
\thanks{Corresponding author: Chang-Xing Ma (cxma@buffalo.edu)}

\begin{abstract}
Measurements are generally collected as unilateral or bilateral data in clinical trials or observational studies. For example, in ophthalmology studies, the primary outcome is often obtained from one eye or both eyes of an individual. In medical studies, the relative risk is usually the parameter of interest and is commonly used. In this article, we develop three confidence intervals for the relative risk for combined unilateral and bilateral correlated data under the equal dependence assumption. The proposed confidence intervals are based on maximum likelihood estimates of parameters derived using the Fisher scoring method. Simulation studies are conducted to evaluate the performance of proposed confidence intervals with respect to the empirical coverage probability, the mean interval width, and the ratio of mesial non-coverage probability to the distal non-coverage probability. We also compare the proposed methods with the confidence interval based on the method of variance estimates recovery and the confidence interval obtained from the modified Poisson regression model with correlated binary data. We recommend the score confidence interval for general applications because it best controls converge probabilities at the 95\% level with reasonable mean interval width. We illustrate the methods with a real-world example. 
\end{abstract}
\maketitle
\section{Introduction}
Binary correlated outcomes are common in clinical trials and research studies in public health and medicine. The correlated data may come from observations of individuals in the same intervention, repeated measurements of a subject at multiple time points, outcomes from members in the same family, responses from paired body parts of the same individual (such as two eyes or two ears of the same individual), etc. 
In studies that involve paired organs, outcomes may be the binary correlated data obtained from both organs of the same individual. However, for various reasons, sometimes we may only be able to collect data on one of the paired organs from an individual\cite{murdoch1998people}, resulting in the so-called combined unilateral and correlated bilateral data \cite{ma2021testing}. For example, in the double-blind randomized clinical trial conducted to compare two antibiotic treatments for the treatment of acute otitis media with effusion \cite{Mandel_1982}, 214 children were randomly assigned to receive either amoxicillin or cefaclor after undergoing unilateral or bilateral tympanocentesis. The study compared children with effusion-free ears and those "improved" from their original status (children with bilateral middle ear effusions at entry but only unilateral middle ear effusions after the treatment) in both treatment groups. Thus, some patients contribute data on two ears while others contribute data on one ear. In such situations, correlations between two ears of the same individual need to be considered. Ignoring the correlation will falsely estimate the variability, resulting in invalid p-values and confidence intervals \cite{murdoch1998people}\cite{sainani2010importance}\cite{ying2018tutorial}\cite{fleiss2013statistical}. Many standard statistical tests that assume independent observations cannot be applied in this situation.

Various methodological research on analyzing binary correlated data has been done that considers the within-cluster correlation, commonly used methods are the generalized linear mixed model (GLMM), and the marginal model using generalized estimating equation (GEE) approaches \cite{ying2018tutorial}\cite{ying2017tutorial}. GLMM incorporates fixed effects and random effects and can be applied to non-normal data such as data from Binomial distributions and Poisson distributions by using appropriate link functions \cite{breslow1993approximate}\cite{ten1999comparison}. To take the intraclass correlation into account, for example, the correlation between ears in the above example, covariance structures of random effects need to be specified. GEE approach was originally proposed to analyze the longitudinal data \cite{zeger1986longitudinal}, it has then been extended to analyze various types of clustered data \cite{liang1993regression}. It is a marginal model-based approach that is very useful when the joint distribution is not available. A working correlation matrix is selected to address the correlation, and a sandwich estimator is used to adjust the variance of the estimator in fitting the marginal model. Methods based on maximum likelihood estimates and tests have also been developed to address the correlation for binary correlated data which can provide an iterative form or an explicit form of the test statistics, for example, homogeneity tests proposed by \citet{Ma_2013R}, \citet{Ma_2013rho}, and \citet{ma2021testing}, common tests and interval estimations on the relative risk, the difference by \citet{Zhuang_2018}, \citet{Zhuang_2018CI}, \citet{shen2017testing}, \citet{xue2019interval}, and \citet{peng2019asymptotic}.

Epidemiology and clinical research is largely based on the measurement of the relative effect \cite{zou2004modified}. When the parameter of interest is the relative risk of an exposure, commonly used methods for independent binary data are the modified Poisson regression model \cite{barros2003alternatives}\cite{zou2004modified} and the log-binomial model \cite{skove1998prevalence}\cite{barros2003alternatives}\cite{wacholder1986binomial}, which have been discussed in many papers \cite{yelland2011performance}\cite{zou2013extension}\cite{chen2018comparing}\cite{petersen2008comparison}\cite{li2021sample}. The log-binomial model uses the log link function for the binomial regression, while the modified Poisson regression applies the Poisson regression to binomial data and uses a robust error variance procedure known as the sandwich estimation to account for the model mis-specification. Both methods use log link functions and thus give a straightforward estimate of the relative risk. The relative risk can then be estimated as the exponential of the coefficient of the indicator. Logistic regression is not a good choice in such situations since the required computations are tedious if we adjust the relative risk from the logistic regression \cite{flanders1987large}\cite{joffe1995standardized}. In addition, naïve conversion of an adjusted odds ratio from the logistic regression to a relative risk has issues such as invalid confidence limits and inconsistent estimates for relative risks\cite{zou2004modified}\cite{mcnutt2003estimating}. The issue of using the log-binomial model is the non-convergence problem \cite{zou2004modified}\cite{petersen2008comparison}\cite{yelland2011performance}. The modified Poisson regression overcomes the non-convergence problem in the log-binomial regression, and is as flexible and powerful as the binomial regression\cite{yelland2011performance}\cite{zou2004modified}, it has become a popular alternative to the log binomial model \cite{zou2013extension}\cite{chen2018comparing}\cite{li2021sample}. When the binary data is clustered, the modified Poisson regression in the GEE framework with the robust sandwich variance estimator can be used to account for clustering effects and model mis-specification \cite{zou2013extension}. In situations where the correlation structure differs among groups, although the modified Poisson regression model with binary correlated data can give consistent estimates, the model may be less efficient \cite{zeger1986longitudinal} due to incorrectly specified working correlation matrix since only one correlation structure can be set for the whole model with the GEE framework. In addition, we may need to consider different working correlation structures to achieve convergence in the real analysis \cite{yelland2011performance}. An alternative of the modified Poisson regression with the robust sandwich variance is a mixed-effect model with random cluster effects, but the distribution of the random effects may be difficult to verify, and the model mis-specification may have a large impact on the results \cite{yelland2011performance}.

In addition to the model-based methods for estimating relative risks, several test-based methods have been proposed. For example, \citet{peng2019asymptotic} proposed several asymptotic confidence intervals (CI) for relative risks for correlated binary data, \citet{xue2019interval}, and \citet{Zhuang_2018CI} further developed CI methods of relative risks for stratified correlated binary data under different assumptions for the intraclass correlation. When the sample size is small, \citet{wang2015exact} derived twelve exact intervals for the relative risk and the odds ratio for data collected from a matched-pairs design or a two-arm independent binomial experiment. However, none of these methods can be applied to the combined unilateral and correlated bilateral data.  

In this paper, we propose three interval estimation methods for combined unilateral and bilateral correlated data. Hypothesis tests of the three proposed methods are also given. We use Rosner’s model to take care of the intraclass correlation that assumes equal dependence across groups and allows different intraclass correlations in different groups. The three parametric methods are based on maximum likelihood estimates (MLE) solved from the real root of the fourth-order polynomial and Fisher iteration methods, which are very efficient and have no convergence issues. We also include the method of the variance estimates recovery (MOVER), which does not consider the intraclass correlation as a baseline method, and the modified Poisson regression for correlated binary data as a comparison of the model-based method. 

In Section 2, we derive the constrained and unconstrained MLEs, propose three CIs, and discuss a baseline method (MOVER) and a model-based method (the modified Poisson regression approach) as comparisons. In Section 3, simulation studies are conducted to evaluate the performance of the proposed methods and compare with the other two methods based on the empirical coverage probability (ECP), the mean interval width (MIW), and the ratio of mesial non-coverage probability to the distal non-coverage probability (RMNCP). Section 4 illustrates the proposed methods with a real dataset. Finally, we give some concluding remarks in Section 5. \\
\section{Methods}
\subsection{Notation and setting}\hfill\\
Without loss of generality, we use notations in the example mentioned above for simplicity. However, the proposed methods can also be applied to other studies with outcomes obtained on paired organs. Suppose there are two groups of combined unilateral and bilateral correlated data (see Table~\ref{tabData}). 
For those who contribute bilateral data to the analysis, let $m_{ti}$ denote the number of patients who have $t$ ($t=0,1,2$) ears with a response (e.g., the presence or absence of some disease or condition) in the $i$th ($i=1,2$) group, $m_{t1}+m_{t2}=S_t$ denote the total number of patients with $t$ ears, and $\sum_{t=0}^2m_{ti}=m_i$ denote the total number of patients in the $ith$ group. 
For patients who contribute unilateral data to the analysis, let $n_{ki}$ denote the number of patients who have $k$ ($k=0,1$) ears with a response in the $i$th group, $n_{k1}+n_{k2}=N_k$ denote the total number of patients with $k$ ears, and $\sum_{k=0}^1n_{ki}=n_i$
denote the total number of patients in the $ith$ group.
Therefore, the total number of patients is $S_0+S_1+S_2=M$ in the bilateral group and $N_0+N_1=N$ in the unilateral group.

To analyze the binary correlated data as shown in Table~\ref{tabData}, the correlation between two ears of the same patient needs to be taken into account. Ignoring the correlation will falsely estimate standard errors and result in invalid p-values and powers \cite{Association_WALTER}\cite{fleiss2013statistical}\cite{fleiss2013statistical}\cite{armstrong2013statistical}. Here, we use the model proposed by Rosner (R model) \cite{Rosner_1982}, which assumes equal dependence between two ears of the same patient in the two groups, to address the intraclass correlation. 
Let $\pi_i$ ($i=1,2$) be the probability of having a response in a ear in the $ith$ group, Rosner's model \cite{Rosner_1982} assumes,
\begin{equation}
Pr(Z_{ijr}=1)=\pi_i, Pr(Z_{ijr}=1|Z_{ij,3-r}=1)=R\pi_i, \label{conditional_probability}
\end{equation}
where $Z_{ijr}=1$ if the $r$th ear of the $j$th patient in the $i$th group has a response at the end of the study, and 0 otherwise, $i=1, 2$, $j=0,\ldots, m_{i}+n_{i}, r=1,2$.  $R$ is a positive constant that measures the dependency between ears of the same patient. Note that $R$ satisfies $0<R\leq 1/a$, if $a\leq 1/2$; $ (2-1/a)/a\leq R\leq 1/a$, if $a > 1/2$, $a=max\{\pi_i, i=1, 2\}$ \cite{Ma_2013R}\cite{ma2021testing}. Based on (\ref{conditional_probability}), it is easy to calculate the correlation between two ears of the same patient in the $ith$ group, which is $\rho_i=corr(Z_{ij1}, Z_{ij2}) = \frac{\pi_i}{1-\pi_i}(R-1)$ \cite{ma2021testing}.

Therefore, for the $i$th group, the observed data $(m_{0i},m_{1i},m_{2i})$ and $n_{1i}$ follow the multinomial and binomial distribution\cite{ma2021testing}, respectively:
$$ (m_{0i},m_{1i},m_{2i})\sim Multinomial(m_i,(R{\pi_i}^2-2\pi_i+1, 2\pi_i(1-R{\pi_i}), R{\pi_i}^2)),$$
$$ n_{1i}\sim Binomial(n_i,\pi_i). $$
Thus, the log-likelihood function is given by
\begin{eqnarray*}
\begin{split}
 l(\pi_1,\pi_2, R) =&  \sum_{i=1}^2 [m_{0i}\log \left(R\, {\pi_i}^2 - 2\, \pi_i + 1\right) +  m_{1i}\log\left(2\pi_i(1-R\pi_i)\right) +  m_{2i}\log\left(R\, \pi_i^2\right)]\\
  +&  \sum_{i=1}^2 [n_{0i}\log \left(1- \pi_i\right) +  n_{1i}\log \pi_i] + constant.
\end{split}
\end{eqnarray*}
\begin{table}
\caption{\label{tabData}Data structure in a combined unilateral and bilateral design.}
\begin{tabular}{cccccc}  \hline
&&\multicolumn{2}{c}{group}\\\cline{3-4}
number of ears with a response &&  1     &  2 && total \\\hline
 0 &&  $m_{01}$   & $m_{02}$  &&$S_{0}$\\
 1 &&  $m_{11}$   & $m_{12}$ &&$S_{1}$\\
 2 &&  $m_{21}$   & $m_{22}$ &&$S_{2}$\\
 total&& $m_{1}$   & $m_{2}$&&$M$ \\ \hline
 0 && $n_{01}$ & $n_{02}$ && $N_{0}$\\
 1 && $n_{11}$ & $n_{12}$ && $N_{1}$\\
 total && $n_{1}$ & $n_{2}$ && $N$\\\hline
\end{tabular}
\end{table}
Let $\delta=\pi_2/\pi_1 $ denote the relative risk we focused on, 
the log-likelihood function can then be expressed as
\begin{eqnarray*}
 l(\delta,\pi_1, R) &=&  m_{01}\log \left(R\, {\pi_1}^2 - 2\, \pi_1 + 1\right) +  m_{11}\log\left(2\pi_1(1-R\pi_1)\right) +  m_{21}\log\left(R\, \pi_1^2\right)\\
  &+&  n_{01}\log \left(1- \pi_1\right) +  n_{11}\log \pi_1\\
   &+& m_{02}\log \left(R\, {(\pi_1\delta)}^2 - 2\, (\pi_1\delta) + 1\right) +  m_{12}\log\left(2(\pi_1\delta)(1-R(\pi_1\delta))\right) +  m_{22}\log\left(R\, (\pi_1\delta)^2\right)\\
  &+&  n_{02}\log \left(1- (\pi_1\delta)\right) +  n_{12}\log (\pi_1\delta)
  + constant.
\end{eqnarray*}
We are interested in testing $H_0:\delta=\delta_{0}$ vs. $H_a:\delta \neq\delta_{0}$.\\
\subsection{ Maximum-likelihood estimates }

\subsubsection{Unconstrained MLEs}\hfill\\ 
 We first derive the unconstrained MLEs of $\pi_i$ and $R$, $i=1,2$, under the alternative hypothesis, which can be solved from 
\begin{equation}
\frac{\partial l}{\partial \pi_i}(\pi_1, \pi_2, R) = \frac{2\, m_{2i}}{\pi_{i}} + \frac{\left(2\, R\, \pi_{i} - 2\right)\, m_{0i}}{R\, {\pi_{i}}^2 - 2\, \pi_{i} + 1} + \frac{\left(4\, R\, \pi_{i} - 2\right)\, m_{1i}}{2\, \pi_{i}\, \left(R\, \pi_{i} - 1\right)} + \frac{n_{1i}}{\pi_i} - \frac{n_{0i}}{1-\pi_i}=0,
\label{eq:deril1}
\end{equation}
and
\begin{equation}
\frac{\partial{l}}{\partial R}(\pi_1, \pi_2, R) =
\frac{S_2}{R} + \sum_{i=1}^g \left[ \frac{{\pi_{i}}^2\, m_{0i}}{R\, {\pi_{i}}^2 - 2\, \pi_{i} + 1} + \frac{\pi_i\, m_{1i}}{R\, \pi_i - 1}\right]=0.
\label{eq:deriR}\end{equation}
There are no closed-form solutions for Equations (\ref{eq:deril1}) and (\ref{eq:deriR}). Following the iteration procedures outlined by \citet{ma2021testing}, $R$ can be updated using the Fisher scoring method while $\pi_i$ can be obtained from the real root solution of the fourth-order polynomial. The iteration steps are repeated until the difference between two estimates of $R$ is sufficiently small. Denote the unconstrained MLEs of $\pi$'s and $R$ by $\hat \pi_i, i=1, 2$, and $\hat R$, respectively. The MLE of $\delta$ can thus be expressed as $\hat \delta=\hat \pi_2/\hat \pi_1$.\\
\subsubsection{Constrained MLEs}\hfill\\ 
Under $H_0$, the relative risk $\pi_2/\pi_1=\delta_0 $, one can solve the following equations to estimate $\pi_1$ and $R$ 
\begin{equation}
{\partial l(\delta_0,\pi_1, R)\over\partial \pi_1}=0,\   {\partial l(\delta_0,\pi_1, R)\over\partial R}=0.
\label{Derivative_R_pi1_under_H_0}
\end{equation}
Again, there are no closed form solutions for \eqref{Derivative_R_pi1_under_H_0}, we use the Fisher scoring algorithm of the form (\ref{Derivative_R_pi1_under_H_0_iterative}) to obtain the constrained MLEs 

\begin{equation}
\begin{pmatrix}  \pi_1^{(t+1)} \\  R^{(t+1)} \end{pmatrix}=\left.\begin{pmatrix} \pi_1^{(t)} \\  R^{(t)}\end{pmatrix}+I^{-1}(\pi_1^{(t)},R^{(t)})\begin{pmatrix}  {\partial l(\delta_0,\pi_1, R)\over\partial \pi_1} \\  {\partial l(\delta_0,\pi_1, R)\over\partial R} \end{pmatrix}\right|_{\pi_1=\pi_1^{(t)}, R=R^{(t)}},
\label{Derivative_R_pi1_under_H_0_iterative}
\end{equation}
where $\pi_1^{(t)}$ and $R^{(t)}$ denote estimates of $\pi_1$ and $R$ from the $t$th iteration, $I^{-1}(\pi_1^{(t)}, R^{(t)})$ is inverse of the Fisher information matrix for ($\pi_1^{(t)}, R^{(t)}$) (see Appendix \ref{appendix_constrainMLE} for the formula of $I(\pi_1, R)$, ${\partial l(\delta_0,\pi_1, R)\over\partial \pi_1}$, and $ {\partial l(\delta_0,\pi_1, R)\over\partial R}$). Here, the unconstrained MLEs $\hat \pi_i, i=1, 2$, and $\hat R$ are set as initial values in the iteration. The iteration stops when the difference of estimates between two steps is sufficiently small (e.g., $|\pi_1^{(t+1)}-\pi_1^{(t)}|<10^{-6}$, $|R^{(t+1)}-R^{(t)}|<10^{-6}$). The constrained MLEs of $\pi_1$ and $ R$ are denoted as $\hat{\pi}_{1H_0}$ and $\hat{R}_{H_0}$, respectively.\\
\subsection{Hypothesis tests and confidence intervals }\hfill\\
In this section, we propose three confidence intervals, the score confidence interval (SC), the profile likelihood confidence interval (PL), and the Wald-type confidence interval (W). Test statistics of the three methods are also provided. The proposed confidence intervals are based on MLEs derived in Section 2.2. In addition, we introduce two existing methods, the method of variance estimates recovery (MOVER) \cite{donner2012closed} and the modified Poisson regression model. The MOVER-based confidence interval does not consider the intraclass correlation and uses the adjusted simple proportion estimates rather than MLEs. Therefore it can be seen as a baseline method. The modified Poisson regression model is a popular method for estimating relative risk for binary data. In the context of correlated binary data, \citet{Zou_2013} extended the modified Poisson regression model by applying the sandwich variance estimator to account for both the cluster effect and the model mis-specification. The modified Poisson regression approach has been proven to have as reliable relative risk estimates as that obtained from the log-binomial regression \cite{Zou_2013}, but does not have convergence problems that are common in the log-binomial regression. We include this model-based method as a comparison of our proposed test-based methods. 
\subsubsection{Score test and score confidence interval (SC)}\hfill\\
The score test statistic to test $H_0:\delta=\delta_{0}$ is given by 
$$T_{SC}^2=\left({\partial l(\delta,\pi_1, R)\over\partial \delta}\right)^2I^{\delta \delta} |  \delta=\delta_0,\pi_1=\hat{\pi}_{1H_0},R=\hat{R}_{H_0}. $$
where $I^{\delta \delta}$ is the upper leftmost entry of the inverse of the Fisher information matrix of $(\delta,\pi_1,R)$ (See Appendix \ref{appendix_derive_scoreCI_WaldCI} for the formula derivation of $I^{\delta \delta}$).
Under the null hypothesis, $T^2_{SC}$ is asymptotically distributed as a chi-square distribution with $1$ degree of freedom. Thus, the $100(1-\alpha)\%$ confidence interval for the relative risk $\delta$ includes values of $\delta_0$ that satisfy the following inequality 
$$
\left\{\delta_0|T_{SC}^2(\delta_0)\leq \chi^2_{1, 1-\alpha}\right\},
$$
where $\chi^2_{1,1-\alpha}$ is the $100(1-\alpha)$th percentile of the chi-square distribution with 1 degree of freedom. The lower and upper bounds of $\delta$ can be solved from the two roots of the equation $T_{SC}^2(\delta_0)=\chi^2_{1, 1-\alpha}$. We use the following steps to obtain the upper bound of $\delta$: \\
Step 1: Set unconstrained MLEs for $(\delta, \pi_1, R)$ as initial values $\hat \delta^{(1)}, \hat \pi_1^{(1)},\hat R^{(1)}$.\\
Step 2: Let $\hat\delta^{(2)}=\hat\delta^{(1)}+flag \times stepsize$, set $flag=1, stepsize=0.1$. Follow the iteration algorithm for obtaining constrained MLEs in Section 2.2.2, we can have MLEs for $\pi_1$ and $R$ under $\delta=\hat\delta^{(2)}$, denote them by $\hat\pi_1^{(2)} $ and $\hat R^{(2)}$.\\
Step 3: if $flag \times \left({\partial l(\delta,\pi_1, R)\over\partial \delta}\right)^2I^{\delta \delta}(\delta,\pi_1,R) |\delta=\hat\delta^{(2)}, \pi_1=\hat\pi_1^{(2)},R=\hat R^{(2)}<flag \times \chi^2_{1,1-\alpha}$, return to Step 2. Otherwise, set $flag=-flag$, $stepsize=0.1 \times stepsize$ and return Step 2.\\
Step 4: Repeat Step 2 and Step 3 until $stepsize< 10^{-5}$. Report $\hat\delta^{(t)}$ as the upper bound of $\delta={\pi_2 \over \pi_1}$.

The lower bound of $\delta$ can be solved by setting $flag=-1$ in Step 2 and changing the condition to $flag \times \left({\partial l(\delta,\pi_1, R)\over\partial \delta}\right)^2I^{\delta \delta}(\delta,\pi_1,R) |\delta=\hat\delta^{(2)}, \pi_1=\hat\pi_1^{(2)},R=\hat R^{(2)}>flag \times \chi^2_{1,1-\alpha}$ in Step 3.\\
\subsubsection{Likelihood ratio test and profile likelihood confidence interval (PL)}\hfill\\
The likelihood ratio test is given by
$$T^2_{LR}=2[l(\hat\delta, \hat\pi_1,\hat R)-l(\delta_0,\hat\pi_{1H_0}, \hat R_{H_0})].$$
Under the null hypothesis $H_0:\delta=\delta_{0}$, $T^2_{LR}$ asymptotically follows a chi-square distribution with 1 degree of freedom. Similar to the score confidence interval in Section 2.3.1, the $100(1-\alpha)\%$ profile likelihood confidence interval for the relative risk can be evaluated by 
$$
\left\{\delta_0|T_{LR}^2(\delta_0)\leq \chi^2_{1, 1-\alpha}\right\},
$$
where $\chi^2_{1,1-\alpha}$ is the $100(1-\alpha)$th percentile of the chi-square distribution with 1 degree of freedom. The confidence interval limits can thus be estimated by solving the equation $T^2_{LR}(\delta_0)=\chi^2_{1,1-\alpha}$. Follow similar iteration steps in Section 2.3.1, except that the inequality in Step 3 is replaced by $flag \times T^2_{LR} |\delta=\hat\delta^{(2)}, \pi_1=\hat\pi_1^{(2)},R=\hat R^{(2)}<flag \times \chi^2_{1-\alpha}$, we can obtain the $100(1-\alpha)\%$ profile likelihood confidence interval for $\delta={\pi_2 \over \pi_1}$.\\
\subsubsection{Wald-type test and Wald-type confidence interval (W)}\hfill\\
Under the null hypothesis $H_0:\delta=\delta_{0}$, the Wald-type test statistics can be expressed as
$$
T^2_W=\frac{(\hat \delta-\delta_0)^2}{Var({\hat \delta})}.
$$
Let $\theta= (\delta,\pi_1,R)^T$, $c=(1,0,0)$, denote the unconstrained MLE of $\theta$ by $\hat \theta= (\hat \delta, \hat \pi_1, \hat R)^T$. Then $Var({\hat \delta})=cI^{-1}(\theta) c^T$, where $I^{-1}(\theta)$ is the inverse of the Fisher information matrix of $\theta$. Denote $I^{\delta \delta}=cI^{-1}(\theta) c^T$ (detailed derivation of $I^{\delta \delta}$ is in Appendix \ref{appendix_derive_scoreCI_WaldCI}), the upper bound and lower bound of the $100(1-\alpha)\%$ Wald-type CI can thus be expressed as
$$
U_w=\hat\delta+z_{1-\alpha/2} \sqrt{I^{\delta \delta}}
$$
and 
$$
L_w=max \left\{0, \hat\delta-z_{1-\alpha/2}\sqrt{I^{\delta \delta}}\right\},
$$
respectively.\\
\subsubsection{MOVER-based confidence interval (MV)}\hfill\\
We review a general approach, method of variance estimates recovery (MOVER), which only requires the availability of confidence limits for constructing the confidence interval of the relative risk proposed by \citet{Zou_2008con}. The adjusted simple proportion estimates can be applied to obtain $\pi_i$. Since this method does not consider the intraclass correlation, we use it to compare with our methods to illustrate the importance of incorporating the intraclass correlation.

Let the confidence interval of $\theta_i$ be $[l_i,u_i], i=1,2$, then we have $l_i=\hat\theta_i-Z_{\alpha/2}\sqrt{Var(\hat\theta_i)}$, $u_i=\hat\theta_i+Z_{\alpha/2}\sqrt{Var(\hat\theta_i)}$. This gives $Var(\hat\theta_1)={(\hat\theta_1-l_1)^2 \over Z^2_{\alpha/2}}$ under $\theta=l_1$,  $Var(\hat\theta_1)={(u_1-\hat\theta_1)^2 \over Z^2_{\alpha/2}}$ under $\theta=u_1$, $Var(\hat\theta_2)={(\hat\theta_2-l_2)^2 \over Z^2_{\alpha/2}}$ under $\theta=l_2$, and $Var(\hat\theta_2)={(u_2-\hat\theta_2)^2 \over Z^2_{\alpha/2}}$ under $\theta=u_2$,

Denote the confidence interval of $\theta_1-\theta_2$ by $[L,U]$, $L$ and $U$ are traditionally given by  $L=\hat\theta_1-\hat\theta_2-Z_{\alpha/2}\sqrt{Var(\hat\theta_1)+Var(\hat\theta_2)}$ and $U=\hat\theta_1-\hat\theta_2+Z_{\alpha/2}\sqrt{Var(\hat\theta_1)+Var(\hat\theta_2)}$. Since $L\approx l_1-u_2$, $U\approx u_1-l_2$, substituting the corresponding variance estimators in the above expressions for L and U, respectively, we have
\begin{equation}
L=\hat\theta_1-\hat\theta_2-\sqrt{(\hat\theta_1-l_1)^2+(u_2-\hat\theta_2)^2},
U=\hat\theta_1-\hat\theta_2+\sqrt{(u_1-\hat\theta_1)^2+(\hat\theta_2-l_2)^2}.
\label{mover_general}
\end{equation}

Let the CI of $\pi_1$ be $(L_1, U_1)$, the CI of $\pi_2$ be $(L_2, U_2)$.
The CI of $log\pi_1$ and $log\pi_2$ can then be expressed as $(logL_1,logU_1)$ and $(logL_2,logU_2)$, respectively. Plugging into \eqref{mover_general}, we have the CI of $log\delta_0$:
\begin{equation}
\begin{split}
  L_{\delta_0} &= log(\hat\pi_2-\hat\pi_1)-\sqrt{(log\hat\pi_2-logL_2)^2+(logU_1-log\hat\pi_1)^2}\\
  &= log{\hat\delta_0}-\sqrt{(log{\hat\pi_2 \over L_2})^2+(log{U_1 \over \hat\pi_1})^2},\\
  U_{\delta_0} &=  log(\hat\pi_2-\hat\pi_1)+\sqrt{(logU_2-log\hat\pi_2)^2+(log\hat\pi_1-logL_1)^2}\\
  &=log{\hat\delta_0}-\sqrt{(log{U_2 \over \hat\pi_2})^2+(log{\hat\pi_1 \over L_1})^2}.
   \label{mover_CI}
\end{split}
\end{equation}
The CI of $\delta_0$ is given by 
\begin{equation}
\begin{split}
 [exp(L_{\delta_0}),exp(U_{\delta_0})].
   \label{mover_CI_2}
\end{split}
\end{equation}

To obtain the estimates of $\pi_i, i=1,2$,  we apply the method proposed by \citet{agresti1998approximate} from which

\begin{eqnarray*}
  [L_i,U_i] &=&  \pi_i\pm Z^2_{1-\alpha/2}\sqrt{\pi_i(1-\pi_i)/\thicktilde n_i},\\
 \pi_i&=&{m_{1i}+2m_{2i}+n_{1i}+Z^2_{1-\alpha/2}/2 \over 2m_i+n_i+Z^2_{1-\alpha/2}},\\
 \thicktilde n_i &=& 2m_i+n_i+Z^2_{1-\alpha/2} .
\end{eqnarray*}
Plugging $\hat{\pi}_i$, $\thicktilde n_i$, $L_i, U_i, i=1,2$ into \eqref{mover_CI} and \eqref{mover_CI_2} gives the upper and lower bound of the MOVER-based Agresti and Coull confidence interval.\\
\subsubsection{Modified Poisson regression model-based confidence interval}\hfill\\

When the parameter of interest is the relative risk, one of the most popular approaches for the binary outcome is the modified Poisson regression model suggested by \citet{zou2004modified}. It is a useful alternative to the log binomial regression and overcomes convergence problems of the log binomial regression approach occurred during the iteration procedure \cite{zou2004modified}\cite{yelland2011performance}\cite{zou2013extension}. This method applies the Poisson distribution to the data with a robust error variance to avoid over-estimated standard errors of the relative risk. Furthermore, the modified Poisson regression approach uses a log link, thus it can estimate relative risks directly rather than the odds ratio estimated by the logistic regression model. In the context of correlated binary data, \citet{zou2013extension} further adjusted the middle term of the sandwich estimator used in the modified Poisson regression model to address the correlation and account for the model mis-specification.

Using the Poisson regression, the model is given by:
$$Pr(y_{ijr}=1)=\pi_{i}=exp(\beta_0+\beta_1x_{ijr}),$$
where $y_{ijr}=1$ if the $r$th ear of the $j$th patient in the $i$th group has a response at the end of the study, and 0 otherwise, $i=1,2$, $j=1, \dots, m_i+n_i$, $r=1,2$. $x_{ijr}$ is the indicator variable of the treatment group, $x_{ijr}=1$ if the patient is in group 2, and 0 otherwise. Thus, the relative risk $\delta=\pi_2/\pi_1$ can be estimated by $exp(\hat\beta_{1})$. The idea of the modified Poisson approach is to apply a robust error variance known as the sandwich variance estimator to correct the overestimated variance for the estimated relative risk due to the model mis-specification \cite{zou2004modified}. \citet{Zou_2013} extended this approach to correlated binary data by modifying the middle term in the sandwich estimator, the adjusted sandwich estimator for binary correlated data is given by 
$$
\hat{Var}(\hat{\beta})=A^{-1}BA^{-1},
$$
where $A=\sum_{i}^{}\sum_{j}^{}\sum_{r}^{}x_{ijr}x_{ijr}^{T}\hat{\pi}_{i}$, $B=\sum_{i}^{}\sum_{j}^{}[\sum_{r}^{}x_{ijr}(y_{ijr}-\hat{\pi}_{i})][\sum_{r}^{}(y_{ijr}-\hat{\pi}_{i})x_{ijr}^{T}]$, which is computed by first grouping the score contributions according to patients. In our case where there is only a single binary exposure, the variance estimate for $\hat\beta_1$ can be simplified as:
\begin{eqnarray*}
  \hat{Var}(\hat{\beta_1}) &=& \hat{Var}({\ln\hat\delta}) \\
  &=& \frac{\sum_{j=1}^{m_1}(y_{1j}-2\hat\pi_1)^2+\sum_{j=1}^{n_1}(y_{1j}-\hat\pi_1)^2}{(\sum_{j=1}^{m_1+n_1}y_{1j})^2}+\frac{\sum_{j=1}^{m_2}(y_{2j}-2\hat\pi_2)^2+\sum_{j=1}^{n_2}(y_{2j}-\hat\pi_2)^2}{(\sum_{j=1}^{m_2+n_2}y_{2j})^2}\\
  &=& \frac{(0-2\pi_1)^2m_{01}+(1-2\pi_1)^2m_{11}+(2-2\pi_1)^2m_{21}+(0-\pi_1)^2n_{01}+(1-\pi_1)^2n_{11}}{(n_{11}+m_{11}+2m_{21})^2}\\
  &+& \frac{(0-2\pi_2)^2m_{02}+(1-2\pi_2)^2m_{12}+(2-2\pi_2)^2m_{22}+(0-\pi_2)^2n_{02}+(1-\pi_2)^2n_{12}}{(n_{12}+m_{12}+2m_{22})^2},
\end{eqnarray*}
where $y_{ij}=\sum_{r}^{}y_{ijr}$ is the number of ears with a response for the $j$th subject in the $i$th group. $\pi_1$ and $\pi_2$ are defined as
\begin{eqnarray*}
 \hat\pi_1 &=& \frac{m_{11}+2m_{21}+n_{11}}{2m_1+n_1}\\
 \hat\pi_2 &=& \frac{m_{12}+2m_{22}+n_{12}}{2m_2+n_2}.
\end{eqnarray*}
Thus the CI of $\delta=\pi_2/\pi_1$ based on the modified Poisson regression model has the form
\begin{equation}
\begin{split}
\exp(\ln\hat{\delta}\mp Z^2_{1-\alpha/2}\sqrt{\hat{Var}(\hat{\beta_1})}),
   \label{GEE_CI}
\end{split}
\end{equation}
where $z_{1-\alpha/2}$ is the $100(1-\alpha/2)$th percentile of the standard normal distribution. Since the modified Poisson regression approach is under the GEE framework, we call \eqref{GEE_CI} the GEE-based CI (GE).\\

\section{Simulation studies}
To investigate the performance of CIs discussed in Section 2, simulation studies are performed to compare the empirical coverage probability (ECP), the mean interval width (MIW), and the ratio of mesial non-coverage probability to the distal non-coverage probability (RMNCP). We set the null $H_0:\delta=\delta_{0}$, the formulas for ECP and MIW of a given CI can be expressed as
$$
ECP=\frac{\sum_{i=1}^{N}I[\delta_0 \in (\delta_L^{(i)},\delta_U^{(i)})]}{N}
$$
and
$$
MIW=\frac{\sum_{i=1}^{N}(\delta_U^{(i)}-\delta_L^{(i)})}{N}
$$
respectively, where $N$ denotes the number of replications, $\delta_U^{(i)}$ and $\delta_L^{(i)}$ denote the upper and lower bound for the $i$th replication, respectively. We use RMNCP to measure the bias of a given CI. It is defined as the left non-coverage probability as a proportion of the total non-coverage probability. The formula is given by
\begin{eqnarray*}
  RMNCP &=& \frac{\sum_{i=1}^{N}I[\delta_0 <\delta_L^{(i)}]/N}{\sum_{i=1}^{N}I[\delta_0 <\delta_L^{(i)}]/N+\sum_{i=1}^{N}I[\delta_0 >\delta_U^{(i)}]/N} \\
  &=&  \frac{\sum_{i=1}^{N}I[\delta_0 <\delta_L^{(i)}]}{\sum_{i=1}^{N}I[\delta_0 \notin (\delta_L^{(i)},\delta_U^{(i)})]}.
\end{eqnarray*}

\subsection{Simulation designs}\hfill\\
First, we consider some properly selected parameter settings. Specifically, we set sample size $m_i=n_i=30,50, 100$ (in the tables and figures for the simulation results, the notation is $m=n=30,50, 100$), $i=1,2$, dependency measurement $R=1,2,3$,  relative risk under the null $\delta_0=1,1.5,2$, and the baseline rate $\pi_1=0.2,0.3$ ($\pi_2=\pi_1\delta_0$). For each parameter setting, 10000 data sets are simulated under the null hypothesis,

we then estimate the three proposed CIs, the Mover-based CI, and the CI based on the modified Poisson regression model (GEE-based CI) under 10000 replications. ECPs, MIWs, and RMNCPs can thus be evaluated based on the formulas shown above. We present results in Table 2-4.

We know that the correlation between two ears for the $i$th group $\rho_i$ is determined by $\rho_i=corr(Z_{ij1}, Z_{ij2}) = \frac{\pi_i}{1-\pi_i}(R-1)$, $i=1,2$. Therefore, larger dependency measurement R gives larger between ear correlation $\rho_i$. For the same parameter settings, we set R from 1 to 3 by 0.1 to cover a wide range of $\rho_i$ to evaluate the performance of ECP and MIW of the methods as the change of between ear correlations. Results are shown in Figure 1-4. 
 
In addition to scenarios with specific parameter settings, to comprehensively evaluate the performance of the five CIs, we compare their ECPs, MIWs, and RMNCPs based on randomly generated parameters. 1000 random parameter settings are generated for $\pi_1, R$ and $\delta_0$ from uniform distributions. For each randomly generated parameter setting, 10000 replications are performed under sample size $m_i=n_i=30,50, 100$, $i=1,2$. The simulation results for ECPs, MIWs, RMNCPs are summarized in the boxplots (Figure 5-7). All the CIs are based on the $5\%$ significance level.

{\small\tabcolsep=2pt
\begin{table}\caption{\label{Table_ECP}ECP based on 10000 simulations under different parameter settings.}
\begin{adjustbox}{center}
\begin{tabular}{ccccccccccccccccccccc}\hline
$\pi_1$&$\delta_0$&$R$&&\multicolumn{5}{c}{$m=n=30$}&&\multicolumn{5}{c}{$m=n=50$}&&\multicolumn{5}{c}{$m=n=100$}\\
\cline{5-9}\cline{11-15}\cline{17-21}
&&&& $W$ & $PL$ & $SC$ &$MV$ & $GE$ &&$W$ & $PL$ & $SC$ &$MV$ & $GE$  &&$W$ & $PL$ & $SC$ &$MV$ & $GE$   \\\hline
0.2 & 1.0 & 1.0 &  & \textbf {93.84} &94.79  & 95.13  & \textbf {96.05} & 95.79  &  & 94.22  & 94.56  & 94.71  & 95.30  & 95.09  &  & 95.08  & 95.05  & 95.12  & 95.33  & 95.15    \\ 
       &  & 2.0 &  & \textbf {93.24} &94.14  & 94.99  & \textbf {93.91} & 95.30  &  & \textbf {93.75} & 94.18  & 94.67  & \textbf {93.22} & 95.00  &  & 94.50  & 94.85  & 95.10  & \textbf {93.19} & 95.08    \\ 
       &  & 3.0 &  & \textbf {93.34} &\textbf {93.86} & 95.43  & \textbf {92.03} & 95.35  &  & \textbf {93.85} & 94.33  & 95.27  & \textbf {91.78} & 95.32  &  & 94.31  & 94.52  & 94.81  & \textbf {91.44} & 95.33    \\ 
       & 1.5 & 1.0 &  & 94.26  &95.00  & 95.26  & 95.53  & 95.45  &  & 94.38  & 94.98  & 94.99  & 95.33  & 95.37  &  & 94.99  & 94.99  & 95.05  & 95.24  & 95.22    \\ 
       &  & 2.0 &  & \textbf {92.70} &\textbf {93.95} & 94.80  & \textbf {92.71} & 94.99  &  & 94.13  & 94.35  & 94.79  & \textbf {92.65} & 95.14  &  & 94.85  & 94.79  & 94.99  & \textbf {92.61} & 94.66    \\ 
       &  & 3.0 &  & \textbf {91.42} &94.41  & 94.91  & \textbf {91.21} & 95.68  &  & \textbf {92.91} & 94.43  & 94.82  & \textbf {90.24} & 95.21  &  & 94.01  & 94.88  & 95.19  & \textbf {90.24} & 94.83    \\ 
       & 2.0 & 1.0 &  & \textbf {93.73} &94.40  & 94.72  & 95.12  & 94.96  &  & 94.42  & 94.76  & 94.82  & 95.27  & 94.99  &  & 94.80  & 94.96  & 94.99  & 95.13  & 94.95    \\ 
       &  & 2.0 &  & \textbf {93.14} &94.84  & 95.28  & \textbf {92.93} & 95.61  &  & 94.17  & 94.99  & 95.21  & \textbf {92.62} & 95.30  &  & 95.02  & 95.20  & 95.44  & \textbf {92.60} & 95.23    \\ 
      0.3 & 1.0 & 1.0 &  & 94.67  &94.93  & 95.21  & 95.61  & 95.45  &  & 94.49  & 94.93  & 95.05  & 95.23  & 95.14  &  & 94.83  & 95.13  & 95.19  & 95.40  & 95.33    \\ 
       &  & 2.0 &  & 94.03  &94.11  & 94.97  & \textbf {92.15} & 95.32  &  & 94.34  & 94.12  & 94.55  & \textbf {91.70} & 95.07  &  & 94.57  & 94.56  & 94.80  & \textbf {92.10} & 95.14     \\ 
       & 1.5 & 1.0 &  & 94.30  &94.60  & 94.93  & 95.17  & 95.03  &  & 94.54  & 94.64  & 94.75  & 94.77  & 94.78  &  & 95.11  & 94.93  & 95.01  & 95.25  & 95.18    \\ 
       &  & 2.0 &  & \textbf {93.05} &94.31  & 94.87  & \textbf {91.02} & 95.25  &  & \textbf {93.57} & 94.62  & 95.12  & \textbf {90.76} & 95.37  &  & 94.57  & 95.19  & 95.24  & \textbf {90.59} & 95.40    \\ 
       & 2.0 & 1.0 &  & 94.88  &94.71  & 94.91  & 95.11  & 95.11  &  & 94.82  & 95.04  & 95.10  & 95.12  & 95.17  &  & 95.09  & 95.23  & 95.23  & 95.19  & 95.29    \\

\hline
\multicolumn{21}{l}{Note: Results in this table are for ECP $\times$ 100. Liberal results (ECP$<$94\%) and conservative results (ECP$>$96\%) }\\
\multicolumn{21}{l}{are in \textbf{bold}.}\\
\hline
\end{tabular}
\end{adjustbox}
\end{table}
}
\hspace{5em}
{\small\tabcolsep=3pt
\begin{table}\caption{\label{Table_MIW}MIW based on 10000 simulations under different parameter settings}
\begin{adjustbox}{center}
\begin{tabular}{ccccccccccccccccccccc}\hline
$\pi_1$&$\delta_0$&$R$&&\multicolumn{5}{c}{$m=n=30$}&&\multicolumn{5}{c}{$m=n=50$}&&\multicolumn{5}{c}{$m=n=100$}\\
\cline{5-9}\cline{11-15}\cline{17-21}
&&&& $W$ & $PL$ & $SC$ &$MV$ & $GE$  &&$W$ & $PL$ & $SC$ &$MV$ & $GE$  &&$W$ & $PL$ & $SC$ &$MV$ & $GE$   \\\hline
0.2 & 1.0 & 1.0 &  & 1.248  &1.364  & 1.325  & 1.366  & 1.331  &  & 0.939  & 0.988  & 0.974  & 0.990  & 0.976  &  & 0.653  & 0.669  & 0.665  & 0.670  & 0.665    \\ 
       &  & 2.0 &  & 1.318  &1.458  & 1.410  & 1.398  & 1.475  &  & 0.985  & 1.044  & 1.027  & 0.998  & 1.067  &  & 0.683  & 0.702  & 0.697  & 0.670  & 0.720    \\ 
       &  & 3.0 &  & 1.232  &1.401  & 1.360  & 1.416  & 1.611  &  & 0.918  & 0.993  & 0.978  & 1.009  & 1.157  &  & 0.631  & 0.657  & 0.652  & 0.671  & 0.773    \\ 
       & 1.5 & 1.0 &  & 1.687  &1.847  & 1.781  & 1.815  & 1.778  &  & 1.256  & 1.322  & 1.297  & 1.309  & 1.294  &  & 0.874  & 0.896  & 0.888  & 0.891  & 0.887    \\ 
       &  & 2.0 &  & 1.781  &1.966  & 1.884  & 1.852  & 1.998  &  & 1.325  & 1.401  & 1.370  & 1.325  & 1.445  &  & 0.916  & 0.942  & 0.932  & 0.896  & 0.982    \\ 
       &  & 3.0 &  & 1.540  &1.779  & 1.699  & 1.856  & 2.186  &  & 1.153  & 1.253  & 1.227  & 1.348  & 1.598  &  & 0.780  & 0.812  & 0.805  & 0.899  & 1.070    \\ 
       & 2.0 & 1.0 &  & 2.098  &2.293  & 2.203  & 2.236  & 2.195  &  & 1.576  & 1.658  & 1.623  & 1.631  & 1.616  &  & 1.086  & 1.113  & 1.102  & 1.104  & 1.100    \\ 
       &  & 2.0 &  & 2.235  &2.448  & 2.346  & 2.289  & 2.510  &  & 1.656  & 1.741  & 1.705  & 1.642  & 1.816  &  & 1.144  & 1.172  & 1.162  & 1.113  & 1.237    \\ 
      0.3 & 1.0 & 1.0 &  & 0.925  &0.981  & 0.970  & 0.973  & 0.963  &  & 0.703  & 0.728  & 0.724  & 0.724  & 0.720  &  & 0.493  & 0.502  & 0.501  & 0.501  & 0.499    \\ 
       &  & 2.0 &  & 0.928  &1.004  & 0.988  & 0.987  & 1.116  &  & 0.702  & 0.736  & 0.730  & 0.731  & 0.829  &  & 0.494  & 0.506  & 0.504  & 0.505  & 0.572    \\ 
       & 1.5 & 1.0 &  & 1.217  &1.292  & 1.272  & 1.263  & 1.255  &  & 0.929  & 0.962  & 0.954  & 0.949  & 0.946  &  & 0.649  & 0.661  & 0.658  & 0.656  & 0.655    \\ 
       &  & 2.0 &  & 1.162  &1.261  & 1.233  & 1.292  & 1.513  &  & 0.868  & 0.909  & 0.898  & 0.954  & 1.119  &  & 0.602  & 0.616  & 0.614  & 0.658  & 0.772    \\ 
       & 2.0 & 1.0 &  & 1.504  &1.595  & 1.564  & 1.548  & 1.540  &  & 1.141  & 1.181  & 1.168  & 1.160  & 1.156  &  & 0.796  & 0.809  & 0.805  & 0.802  & 0.801    \\

\hline

\end{tabular}
\end{adjustbox}
\end{table}
}

{\small\tabcolsep=2pt
\begin{table}\caption{\label{Table_RMNCP}RMNCP based on 10000 simulations under different parameter settings.}
\begin{adjustbox}{center}
\begin{tabular}{ccccccccccccccccccccc}\hline
$\pi_1$&$\delta_0$&$R$&&\multicolumn{5}{c}{$m=n=30$}&&\multicolumn{5}{c}{$m=n=50$}&&\multicolumn{5}{c}{$m=n=100$}\\
\cline{5-9}\cline{11-15}\cline{17-21}
&&&& $W$ & $PL$ & $SC$ &$MV$ & $GE$  &&$W$ & $PL$ & $SC$ &$MV$ & $GE$   &&$W$ & $PL$ & $SC$ &$MV$ & $GE$   \\\hline
 0.2 & 1.0 & 1.0 &  & 0.005  &0.508  & 0.507  & 0.532  & 0.506  &  & 0.059  & 0.490  & 0.482  & 0.496  & 0.489  &  & 0.167  & 0.525  & 0.525  & 0.514  & 0.528    \\ 
       &  & 2.0 &  & 0.006  &0.497  & 0.499  & 0.504  & 0.494  &  & 0.037  & 0.465  & 0.469  & 0.488  & 0.470  &  & 0.135  & 0.491  & 0.488  & 0.479  & 0.470    \\ 
       &  & 3.0 &  & 0.002  &0.502  & 0.514  & 0.493  & 0.495  &  & 0.037  & 0.482  & 0.507  & 0.468  & 0.453  &  & 0.149  & 0.467  & 0.457  & 0.484  & 0.465    \\ 
       & 1.5 & 1.0 &  & 0.002  &0.532  & 0.492  & 0.468  & 0.462  &  & 0.037  & 0.468  & 0.439  & 0.415  & 0.400  &  & 0.148  & 0.501  & 0.477  & 0.462  & 0.464    \\ 
       &  & 2.0 &  & 0.003  &0.511  & 0.438  & 0.455  & 0.471  &  & 0.043  & 0.522  & 0.468  & 0.467  & 0.465  &  & 0.171  & 0.534  & 0.479  & 0.465  & 0.474    \\ 
       &  & 3.0 &  & 0.000  &0.503  & 0.412  & 0.424  & 0.449  &  & 0.031  & 0.544  & 0.471  & 0.477  & 0.474  &  & 0.142  & 0.547  & 0.470  & 0.477  & 0.516    \\ 
       & 2.0 & 1.0 &  & 0.000  &0.486  & 0.415  & 0.377  & 0.355  &  & 0.029  & 0.513  & 0.442  & 0.406  & 0.391  &  & 0.152  & 0.486  & 0.449  & 0.427  & 0.426    \\ 
       &  & 2.0 &  & 0.000  &0.526  & 0.390  & 0.431  & 0.444  &  & 0.029  & 0.509  & 0.399  & 0.434  & 0.417  &  & 0.133  & 0.517  & 0.414  & 0.439  & 0.447    \\ 
      0.3 & 1.0 & 1.0 &  & 0.053  &0.511  & 0.520  & 0.517  & 0.523  &  & 0.118  & 0.482  & 0.485  & 0.480  & 0.481  &  & 0.203  & 0.487  & 0.491  & 0.489  & 0.486    \\ 
       &  & 2.0 &  & 0.032  &0.520  & 0.527  & 0.527  & 0.528  &  & 0.108  & 0.513  & 0.508  & 0.504  & 0.513  &  & 0.298  & 0.561  & 0.565  & 0.520  & 0.541      \\ 
       & 1.5 & 1.0 &  & 0.030  &0.517  & 0.477  & 0.451  & 0.443  &  & 0.125  & 0.491  & 0.467  & 0.432  & 0.425  &  & 0.249  & 0.525  & 0.503  & 0.486  & 0.483    \\ 
       &  & 2.0 &  & 0.035  &0.521  & 0.462  & 0.486  & 0.480  &  & 0.078  & 0.510  & 0.445  & 0.449  & 0.456  &  & 0.169  & 0.499  & 0.456  & 0.492  & 0.500    \\ 
       & 2.0 & 1.0 &  & 0.025  &0.510  & 0.473  & 0.387  & 0.378  &  & 0.129  & 0.514  & 0.469  & 0.426  & 0.412  &  & 0.206  & 0.524  & 0.491  & 0.443  & 0.442    \\ 
 
\hline

\end{tabular}
\end{adjustbox}
\end{table}
}

\begin{figure}[htpb]
\centering
\noindent\makebox[\textwidth][c] {
   \includegraphics[width=1\paperwidth]{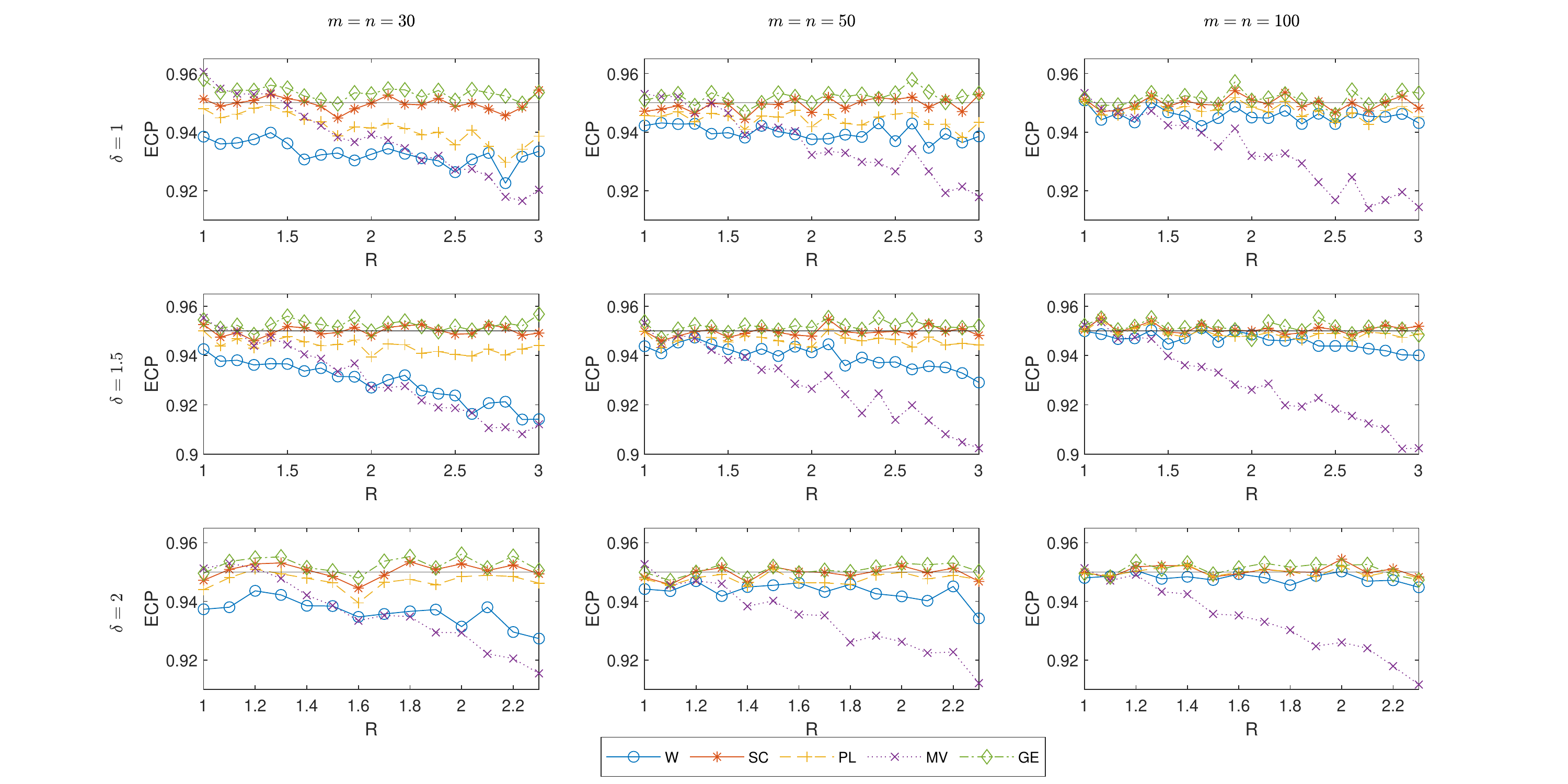}
   }
 \caption{ECP on different dependency measurements R with $\pi_1=0.2$.}

\end{figure}

\begin{figure}[htpb]
\centering
\noindent\makebox[\textwidth][c] {
   \includegraphics[width=1\paperwidth]{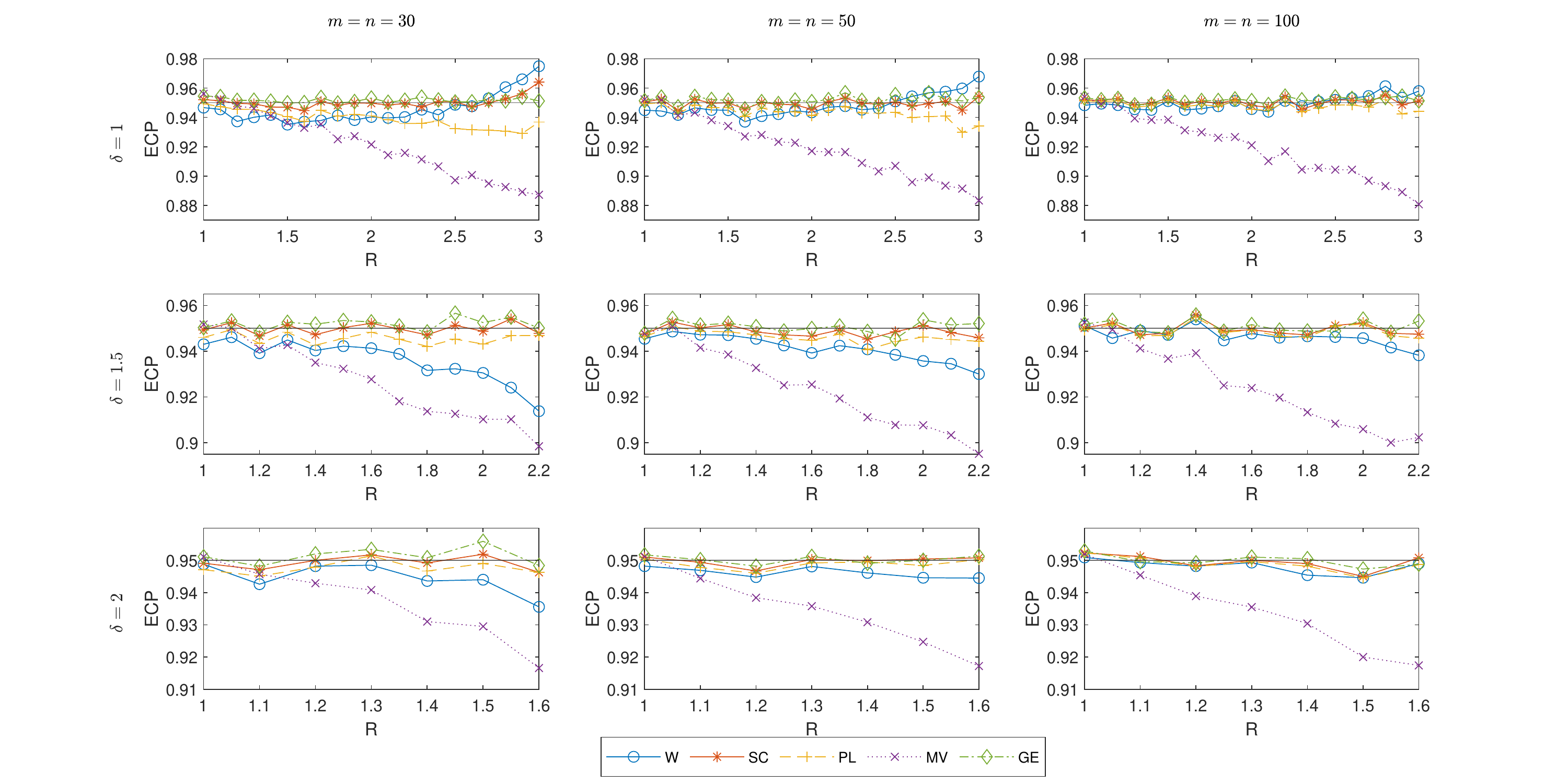}
   }
 \caption{ECP on different dependency measurements R with $\pi_1=0.3$.}

\end{figure}

\begin{figure}[htpb]
\centering
\noindent\makebox[\textwidth][c] {
   \includegraphics[width=1\paperwidth]{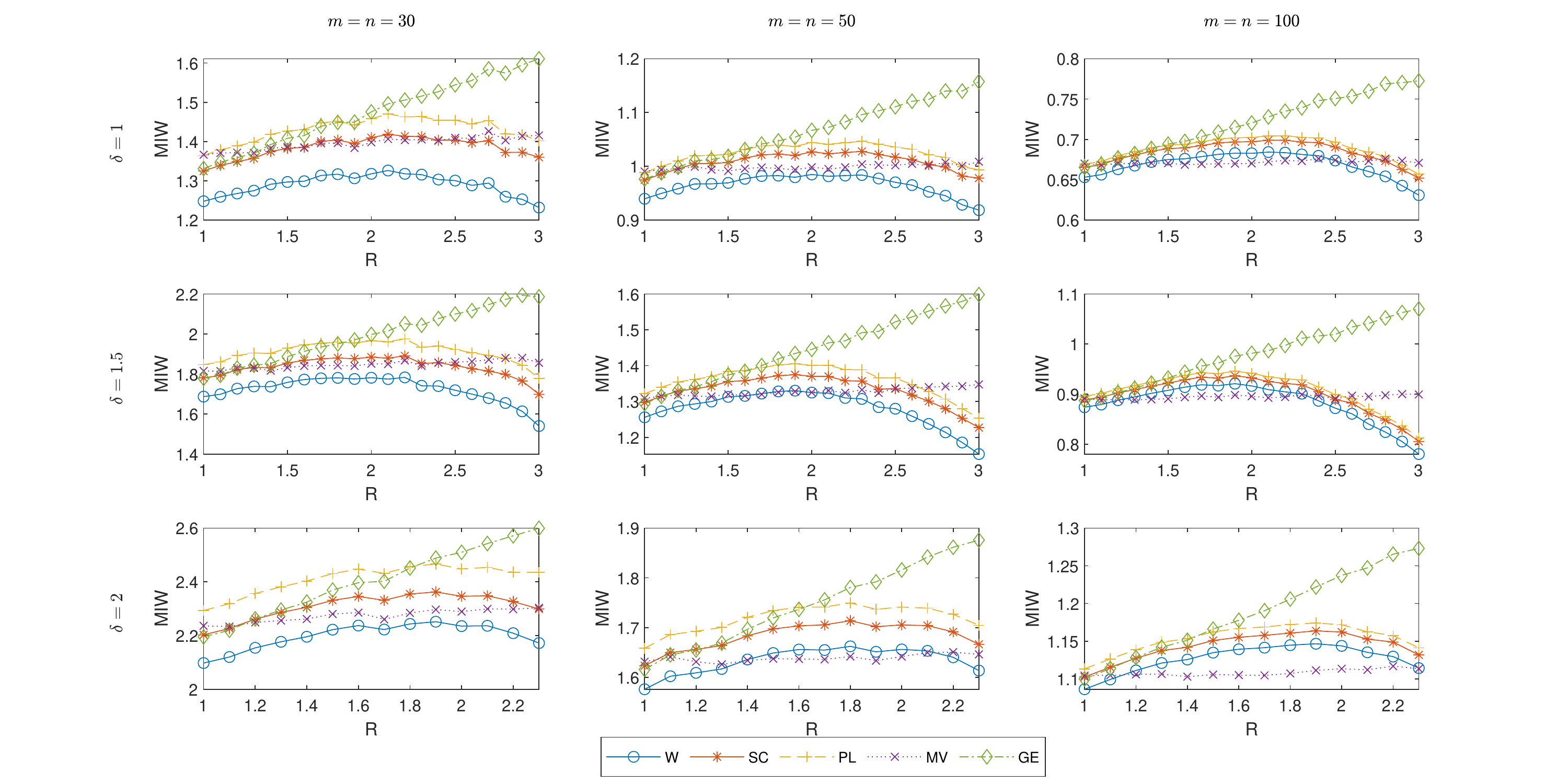}
   }
 \caption{MIW on different dependency measurements R with $\pi_1=0.2$.}

\end{figure}

\begin{figure}[htpb]
\centering
\noindent\makebox[\textwidth][c] {
   \includegraphics[width=1\paperwidth]{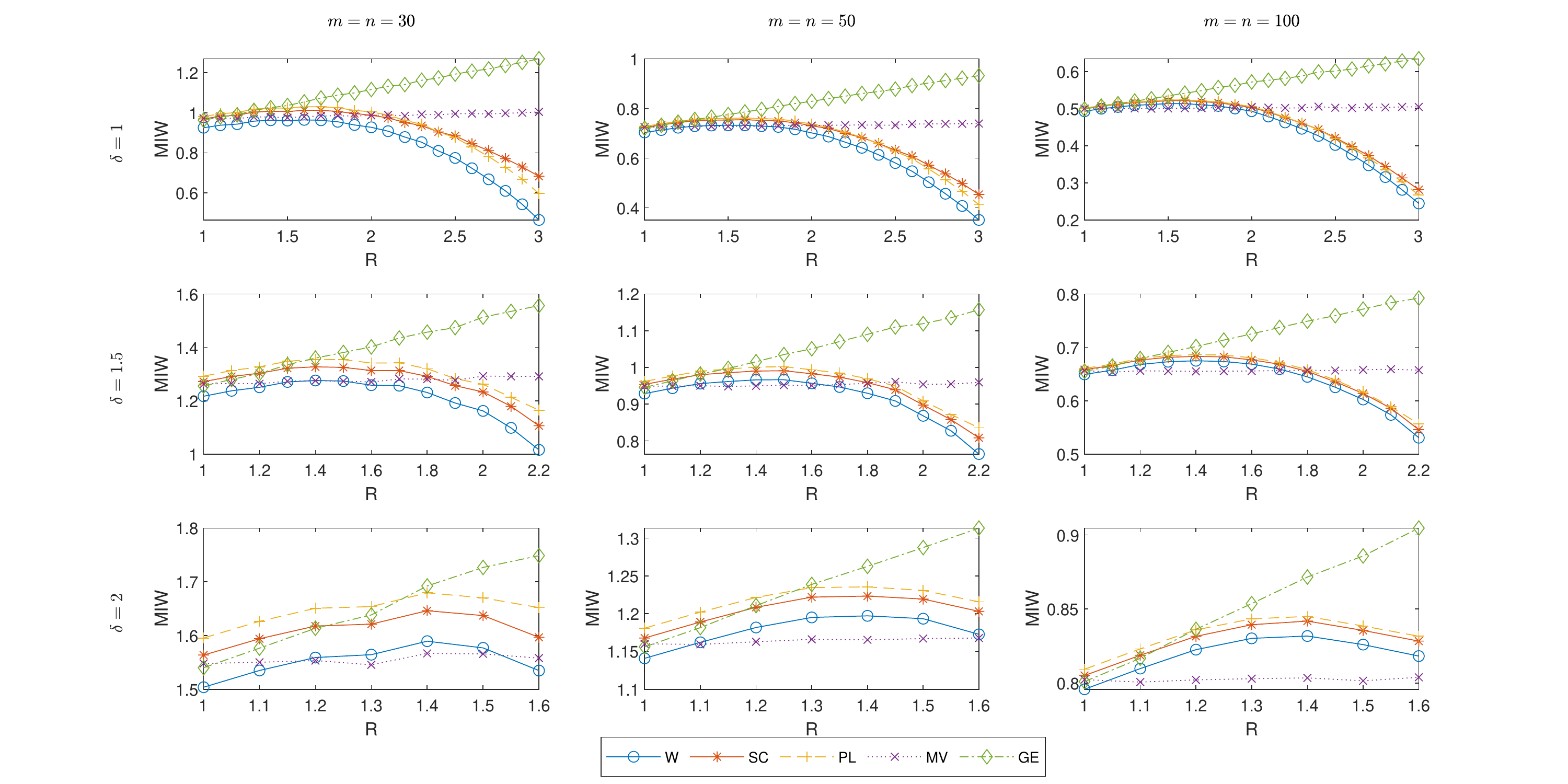}
   }
 \caption{MIW on different dependency measurements R with $\pi_1=0.3$.}

\end{figure}

\begin{figure}[htpb]
\centering
\noindent\makebox[\textwidth][c] {
   \includegraphics[width=1\paperwidth]{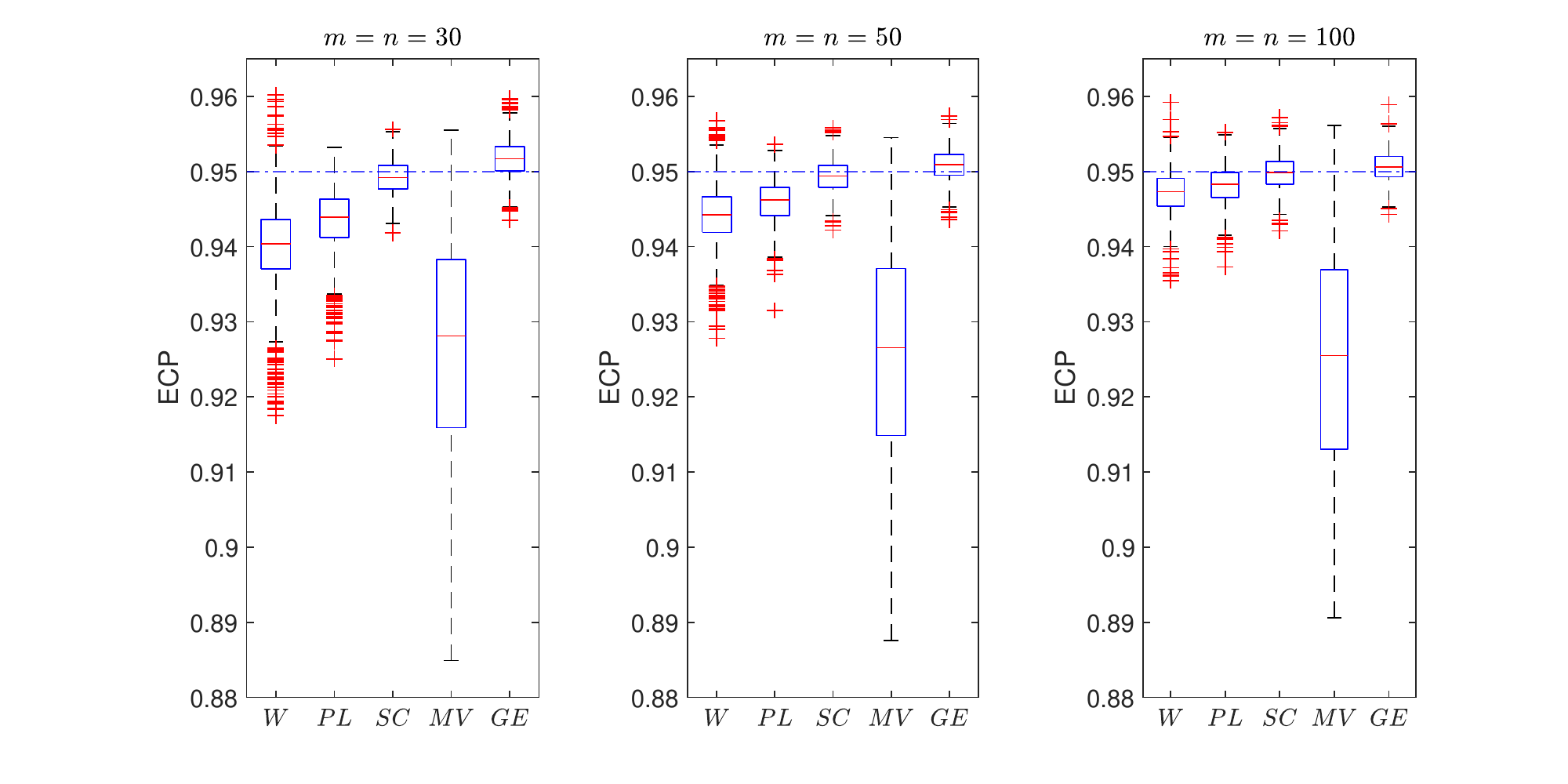}
   }
 \caption{Boxplots for ECP under different sample sizes. The horizontal dashed and dotted line corresponds to the nominal coverage probability.}

\end{figure}

\begin{figure}[htpb]
\centering
\noindent\makebox[\textwidth][c] {
   \includegraphics[width=1\paperwidth]{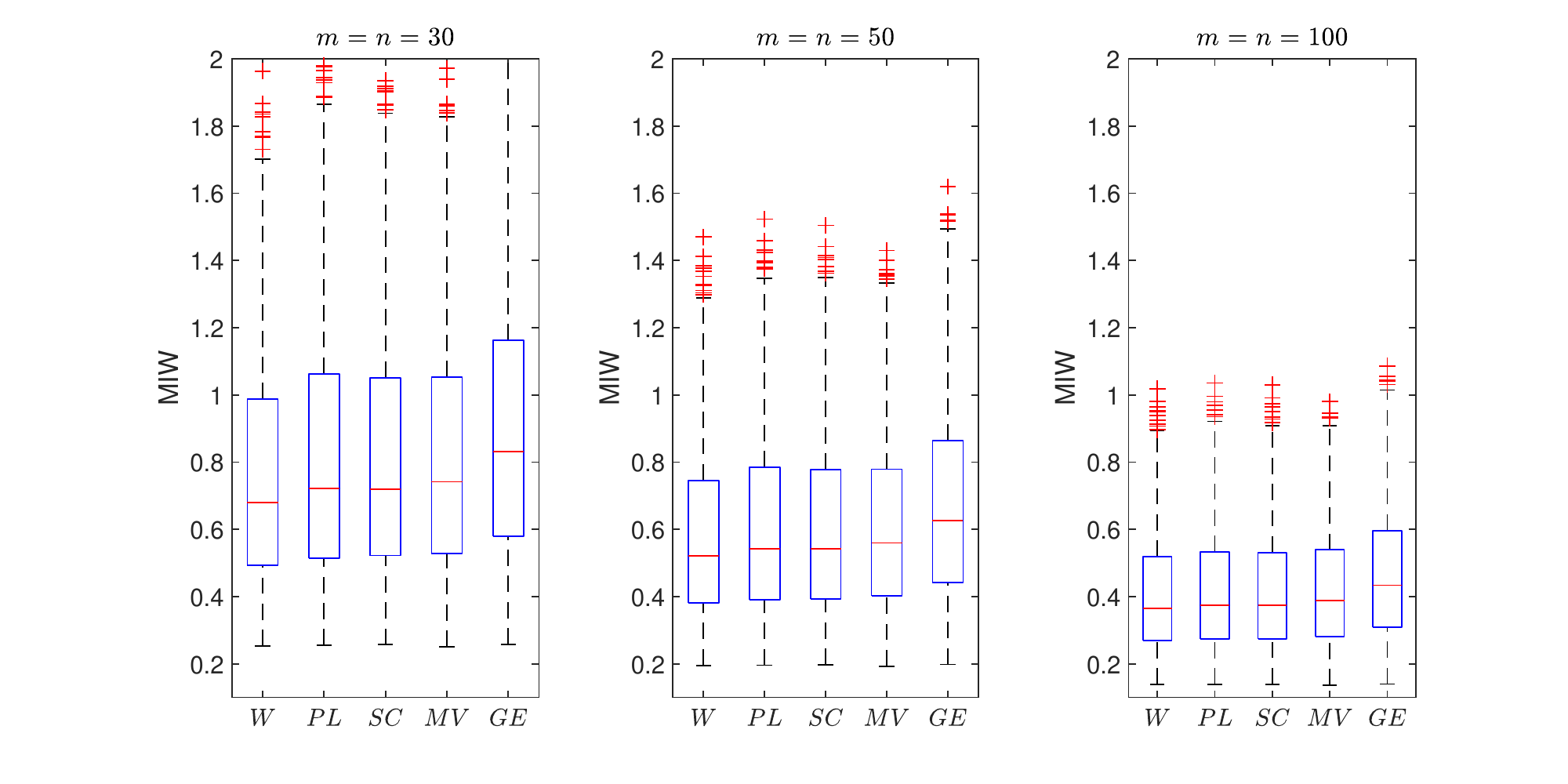}
   } \caption{Boxplots for MIW under different sample sizes.}

\end{figure}

\begin{figure}[htpb]
\centering
\noindent\makebox[\textwidth][c] {
   \includegraphics[width=1\paperwidth]{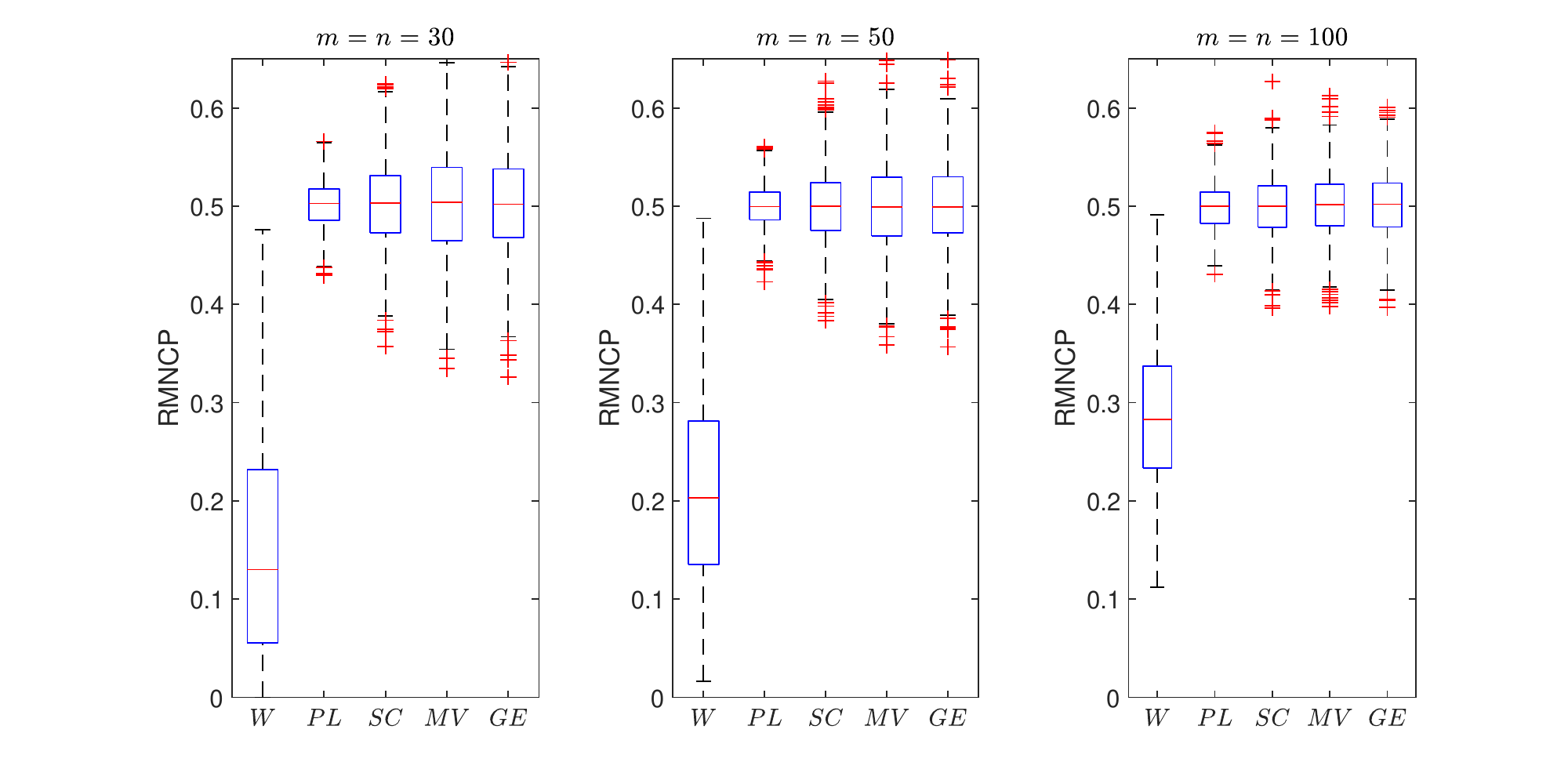}
   } \caption{Boxplots for RMNCP under different sample sizes.}

\end{figure}

\subsection{Simulation results}\hfill\\
We first describe our findings on specific parameter settings (results are shown in Table 2-4). Table 2 summarizes the simulation results for ECPs under specific parameter settings. It is clear that the ECP for the score CI and the GEE-based CI are closer to the nominal coverage in almost all cases, while the Wald-type CI, the profile likelihood CI and the MOVER-based CI have lower coverage in many settings. For MIW (see Table 2), GEE-based CI gives a larger MIW than the other CIs. The simulation results for RMNCP are presented in Table 4. We find that the RMNCP for all the CI methods are close to 0.5 except for the Wald-type CI, indicating that all the methods except for the Wald-type CI have symmetric non-coverage probabilities.

In addition to the specific parameter settings, we investigate the relationship between ECP, MIW, and the intraclass correlation (results are shown in Figure 1-4). The results correspond to the results presented in Table 2-3, but allow us to see the trend of ECP and MIW as the increase of the intraclass correlation. Figure 1 and Figure 2 show the relationship between ECP and R on baseline rate $\pi_1=0.2$ and $\pi_1=0.3$, respectively. Each row in these two figures indicates different relative risks under the null, while each column indicates different sample size scenarios. The score CI performs the best since its ECP is the closest to the nominal coverage in all cases. The GEE-based CI also performs satisfactorily but is slightly conservative compared to the score CI, especially when the sample size is not large (e.g., $m=n=30$). Both the Wald-type CI and the profile likelihood CI have some results well below the nominal coverage probability, while the Wald-type CI is always more liberal than the profile likelihood CI. We also find that the Wald-type CI is closer to the nominal coverage generally under $m=n=30, 50$ for $\pi_1=0.3$ compared to when $\pi_1=0.2$. In addition, when $\delta \neq 1$, the ECP of the Wald-type CI has a decreasing trend when R increases (i.e., the correlation between ears increases). The MOVER-based CI is extremely liberal as expected since the intraclass correlation is not considered and its performance declines quickly as the increase of R. As the sample size goes large, all of the CIs get closer to the nominal coverage except for the MOVER-based CI since all of the methods are asymptotic methods except for the Mover-based CI. Table 3 shows the simulation results for MIW of the five CIs. All the CIs are close to each other except for the GEE-based CI, which gives larger MIWs than the other CIs, especially when R becomes large. From Figure 3-4, we note a monotonically increasing relationship between the GEE-based CI and R. Therefore, although the ECP of the GEE-based CI is close to the nominal coverage probability under all scenarios, this CI is not recommended due to its larger MIW. The proposed three methods give similar MIW while the MIW for the score CI is always between the MIW of the Wald-type CI and the profile likelihood CI.

The boxplots (Figure 5-7) summarize scenarios of randomly generated parameter settings. Even under a larger sample size ($m=n=50, 100$), the ECP (Figure 5) for the score CI and the GEE-based CI are closer to the nominal level than the profile likelihood CI and the Wald CI. Furthermore, as the sample size goes large, MIWs (Figure 6) become smaller for all CIs. Boxplots also confirm that the score CI has the best performance because it produces the closest nominal coverage probability with reasonable MIW and RMNCP on the whole parameter space. Therefore, we recommend the score CI in applications.

\section{Example}

In this section, we use an example to illustrate the interval estimation methods discussed in Section 2. A double-blind randomized clinical trial was conducted at two sites to compare two antibiotics for the treatment of acute otitis media with effusion (OME) \cite{Mandel_1982}. A total of 214 children aged 2 months - 16 years were randomized to receive either amoxicillin or cefaclor after undergoing unilateral or bilateral tympanocentesis. The primary outcome is the effusion-free status at follow-up, measured at the patient level. Table~\ref{tabExmDatafor_us} shows the presence or absence of OME at 42 days, which consists of 173 children out of the sample of 214 children with 93 children in the Cefaclor group and 80 children in the Amoxicillin group. We are interested in the relative risk and the confidence interval of the cured ears in the two treatment groups. In the original study \cite{Mandel_1982}, the conclusion was given by comparing the percentage of children without effusion or "improved" (means those with bilateral middle ear effusions at baseline but only unilateral after the treatment) in the two treatment groups (68.9\% vs. 67.5\%), indicating a relative risk of 0.9797 without considering intraclass correlations. Here, we apply the three proposed methods and the two existing methods to this example to illustrate our methods and compare them to the results in the original study \cite{Mandel_1982}.

The tests and interval estimators developed in this article are based on Rosner's equal dependence model. Thus, goodness of fit tests need to be performed to check if the R model is appropriate for this dataset before using the proposed methods. Several goodness-of-fit tests have been proposed by \citet{liu2020goodness} to examine the intraclass correlation problem arising in bilateral data. Based on the goodness-of-fit tests they developed, both the likelihood ratio test ($G^2$) and the Pearson chi-square test ($\chi^2$) show that the Rosner's model fits for the data well, which give us $G^2=0.3871$ with p-value= 0.5338 and $\chi^2= 0.3867$ with p-value= 0.5341. Therefore, we can apply our methods to this example.

Based on the maximum likelihood estimate algorithm derived in Section 2.2, we obtain the unconstrained MLEs for the probability of children with effusion-free ears and "improved" ears in the Cefaclor group $\pi_1$, the proportion ratio of children with effusion-free ears and "improved" ears after treatments in the two groups $\delta$, and the dependency measurement $R$, which give us $\hat \pi_1=0.6528, \hat\delta=0.9841, \hat R=1.3172$, respectively, intraclass correlations in the two groups are $\hat\rho_1=0.5964$ and $\hat\rho_2=0.5699$. The relative risk estimate based on the maximum likelihood derived in this paper is 0.9841, which is close to the relative risk estimate in the original paper (0.9797). We also obtain estimates of the relative risk and the confidence interval using the proposed three methods and the two methods mentioned in Section 2.3.4 and Section 2.3.5. The results are shown in Table~\ref{tabExmStat_CI}. We see that the MOVER-based CI and the GEE-based CI have smaller relative risk estimates than the proposed three methods. The interval width for the three proposed methods ranges from 0.3123 to 0.3260. The interval width for the GEE-based CI is the largest, which is 0.4217, corresponding to the findings in the simulation studies in Section 3.2 that the GEE-based CI always gives a larger interval width. We can also see that all the CIs include 1, indicating that we fail to reject the null hypothesis $H_0: \delta=\pi_2/\pi_1=1$. It suggests that the probability of ears being cured under the two treatment methods is not significantly different. This result corresponds to the result in the original paper that by 42 days after entry, the proportion of children with effusion-free ears and "improved" ears in the two treatment groups were not different.   
\begin{table*}
\caption{\label{tabExmDatafor_us}Distribution of the number of ears without disease at 42 Days}
\begin{threeparttable}
\begin{tabular}{cccc}  \hline
&&\multicolumn{2}{c}{Treatment Group}\\\cline{3-4} 
Number of ears being cured  & &  Cefaclor     &  Amoxicillin  \\\hline
 0 &&  9  &  7\\
 1 &&  7   &  5 \\
 2 &&   23   &  13 \\
 Total &&   39   &  25 \\\hline 
 0 &&  20   & 19 \\
 1 &&   34   & 36 \\
 Total &&   54   & 55 \\\hline
 
 \end{tabular}
 
      \end{threeparttable}
\end{table*}

\begin{table}
\caption{\label{tabExmStat_CI}Relative risks and the corresponding CIs of proportions of ears being cured in the two treatment groups}
\begin{threeparttable}
\begin{tabular}{lccccc}\hline
Method & $RR^{\star}$ && 95\% CI&& Width  \\\hline
Score CI & 0.9841  && 0.8251-1.1510 &&0.3260\\
Profile likelihood CI &0.9841 &&0.8274-1.1517&&0.3242\\
Wald-type CI & 0.9841  &&0.8280-1.1403 &&0.3123\\
MOVER-based CI &0.9674&&0.7979-1.1658&&0.3680\\
GEE-based CI & 0.9681 &&0.7800-1.2017&&0.4217\\\hline
\end{tabular}
      \small
      \end{threeparttable}
\end{table}

\section{Discussion}
In this article, we developed three confidence intervals for the relative risk: the score CI, the Wald-type CI, and the profile likelihood CI based on MLEs derived using the Fisher scoring method. The Wald-type CI can be easily derived while the score CI and the profile likelihood CI are computed by our efficient searching algorithm. We used Rosner's model, which assumes equal dependence between two ears of the same patient across groups, to address the between ear correlation. The three proposed methods were compared with two existing methods, the MOVER-based method and the modified Poisson regression model. 

Simulation results indicate that the score CI works the best because its empirical coverage probability is the closest to the nominal coverage probability with reasonable mean interval width and RMNCP. The profile likelihood CI and the Wald-type CI are liberal in the simulations, especially when the sample size is relatively small. The MOVER-based CI is extremely liberal because the intraclass correlation is not taken into account. The GEE-based CI is slightly conservative when the sample size is small. For a large sample size, its performance of ECP is very comparable to that of the score CI. However, the GEE-based CI tends to have larger interval widths compared with other methods under large R. Interestingly, there is an increasing trend for the mean interval width of the GEE CI when R becomes large, while the trend for the mean interval width of the three proposed methods is decreasing, which indicates that our methods have a more reliable interval width in a wide range of the intraclass correlation. Therefore, the score CI is recommended for general analysis in that its performance is the best in all settings evaluated.

Model-based methods are more flexible because they can incorporate covariates and allow flexible choices of the cluster size. However, they cannot specify the group-specified correlation structure that addresses different intraclass correlations in different groups as the dataset discussed in this article, and thus may suffer from model mis-specification. In addition, model-based methods cannot provide an explicit form or the iterative form of the test statistics, and the convergence may not be achieved \cite{chen2018comparing}\cite{Zou_2013}\cite{petersen2008comparison}. The three proposed methods and the model-based method are asymptotic methods. Therefore they may not perform well when the sample size is small. With the explicit form of the test statistics as provided in this article, we are able to derive the exact method for small sample size scenarios in the future. Since the proposed methods assume Rosner's model, we should perform goodness of fit tests to evaluate if the data fits the R model before using our methods. Care should also be taken when the true rate is very large or small (close to 0 or 1) or when the intraclass correlation is very large.

\medskip

\bibliographystyle{unsrtnat}
\bibliography{main}

\appendix
\section{ Formula derivation}
\subsection{Information matrix and formula derivation for computing constrained MLEs}\label{appendix_constrainMLE}\hfill\\
The first-order derivative of the log-likelihood with respect to $\pi_1$ and $R$ yield
\begin{eqnarray*}
{\partial l(\pi_1,\delta_0;  R)\over\partial \pi_1}&=& \frac{2S_2+N_1}{\pi _{1}}+\frac{n_{01}}{\pi _{1}-1}+\frac{\delta _{0}\,n_{02}}{\delta _{0}\,\pi _{1}-1}+\frac{m_{01}\,\left(2\,R\,\pi _{1}-2\right)}{R\,\pi _{1}^2-2\,\pi _{1}+1}-\frac{m_{12}\,\left(2\,R\,\delta _{0}\,\pi _{1}-1\right)}{\pi _{1}-R\,\delta _{0}\,\pi _{1}^2}-\frac{m_{11}\,\left(2\,R\,\pi _{1}-1\right)}{\pi _{1}-R\,\pi _{1}^2}\\&+&\frac{2\,\delta _{0}\,m_{02}\,\left(R\,\delta _{0}\,\pi _{1}-1\right)}{R\,{\delta _{0}}^2\,\pi _{1}^2-2\,\delta _{0}\,\pi _{1}+1},\\
{\partial l(\pi_1,\delta_0;  R)\over\partial R}&=&     \frac{S_2}{R}+\frac{m_{11}\,\pi _{1}}{R\,\pi _{1}-1}+\frac{m_{01}\,\pi _{1}^2}{R\,\pi _{1}^2-2\,\pi _{1}+1}+\frac{{\delta _{0}}^2\,m_{02}\,\pi _{1}^2}{R\,{\delta _{0}}^2\,\pi _{1}^2-2\,\delta _{0}\,\pi _{1}+1}+\frac{\delta _{0}\,m_{12}\,\pi _{1}}{R\,\delta _{0}\,\pi _{1}-1}.
\end{eqnarray*}

The second-order derivative of the log-likelihood with respect to $\pi_1$ and $R$ yield
\begin{eqnarray*}
  {\partial^2 l\over\partial \pi_1^2} &=&  \frac{m_{12}\,\left(2\,R\,\delta _{0}\,\pi _{1}-1\right)}{\pi _{1}^2-R\,\delta _{0}\,\pi _{1}^3}-\frac{2\,S_2}{\pi _{1}^2}-\frac{N_1}{\pi _{1}^2}-\frac{n_{01}}{{\left(\pi _{1}-1\right)}^2}-\frac{4\,m_{02}\,{\left(\delta _{0}-R\,\delta _{0}^2\,\pi _{1}\right)}^2}{{\left(R\,\delta _{0}^2\,\pi _{1}^2-2\,\delta _{0}\,\pi _{1}+1\right)}^2}+\frac{2\,R\,m_{01}}{R\,\pi _{1}^2-2\,\pi _{1}+1}\\&-&\frac{\delta _{0}^2\,n_{02}}{{\left(\delta _{0}\,\pi _{1}-1\right)}^2}-\frac{4\,m_{01}\,{\left(R\,\pi _{1}-1\right)}^2}{{\left(R\,\pi _{1}^2-2\,\pi _{1}+1\right)}^2}+\frac{2\,R\,\delta _{0}^2\,m_{02}}{R\,\delta _{0}^2\,\pi _{1}^2-2\,\delta _{0}\,\pi _{1}+1}+\frac{2\,R\,m_{11}}{\pi _{1}\,\left(R\,\pi _{1}-1\right)}-\frac{m_{11}\,\left(2\,R\,\pi _{1}-1\right)}{\pi _{1}^2\,\left(R\,\pi _{1}-1\right)}\\&-&\frac{R\,m_{11}\,\left(4\,R\,\pi _{1}-2\right)}{2\,\pi _{1}\,{\left(R\,\pi _{1}-1\right)}^2}+\frac{2\,R\,\delta _{0}\,m_{12}}{\pi _{1}\,\left(R\,\delta _{0}\,\pi _{1}-1\right)}-\frac{R\,\delta _{0}\,m_{12}\,\left(2\,R\,\delta _{0}\,\pi _{1}-1\right)}{\pi _{1}\,{\left(R\,\delta _{0}\,\pi _{1}-1\right)}^2}, \\
  {\partial^2 l\over\partial \pi_1\partial R} &=&     \frac{m_{11}}{R\,\pi _{1}-1}+\frac{2\,m_{01}\,\pi _{1}}{R\,\pi _{1}^2-2\,\pi _{1}+1}+\frac{\delta _{0}\,m_{12}}{R\,\delta _{0}\,\pi _{1}-1}-\frac{m_{01}\,\pi _{1}^2\,\left(2\,R\,\pi _{1}-2\right)}{{\left(R\,\pi _{1}^2-2\,\pi _{1}+1\right)}^2}+\frac{2\,\delta _{0}^2\,m_{02}\,\pi _{1}}{R\,\delta _{0}^2\,\pi _{1}^2-2\,\delta _{0}\,\pi _{1}+1}\\&-&\frac{R\,m_{11}\,\pi _{1}}{{\left(R\,\pi _{1}-1\right)}^2}-\frac{2\,\delta _{0}^3\,m_{02}\,\pi _{1}^2\,\left(R\,\delta _{0}\,\pi _{1}-1\right)}{{\left(R\,\delta _{0}^2\,\pi _{1}^2-2\,\delta _{0}\,\pi _{1}+1\right)}^2}-\frac{R\,\delta _{0}^2\,m_{12}\,\pi _{1}}{{\left(R\,\delta _{0}\,\pi _{1}-1\right)}^2}, \\
{\partial^2 l\over\partial R^2} &=& -\frac{S_2}{R^2}-\frac{m_{11}\,\pi _{1}^2}{{\left(R\,\pi _{1}-1\right)}^2}-\frac{m_{01}\,\pi _{1}^4}{{\left(R\,\pi _{1}^2-2\,\pi _{1}+1\right)}^2}-\frac{\delta _{0}^4\,m_{02}\,\pi _{1}^4}{{\left(R\,\delta _{0}^2\,\pi _{1}^2-2\,\delta _{0}\,\pi _{1}+1\right)}^2}-\frac{\delta _{0}^2\,m_{12}\,\pi _{1}^2}{{\left(R\,\delta _{0}\,\pi _{1}-1\right)}^2}.
\end{eqnarray*}
Then the information matrix for $\pi_1$ and $R$ is given by
$$I(\pi_1, R)= 
\begin{bmatrix}
  I_{\pi_1,\pi_1} & I_{\pi_1,R} \\
   I_{\pi_1,R} & I_{R,R} \\
\end{bmatrix}
,$$
where
\begin{eqnarray*}  
  I_{\pi_1,\pi_1}&=&E\left(-{\partial^2 l\over\partial \pi_1^2}\right) =4\,R\,m_{1}+\frac{n_{1}}{\pi _{1}}-\frac{n_{1}}{\pi _{1}-1}+\frac{\delta _{0}\,n_{2}}{\pi _{1}}+\frac{4\,m_{2}\,{\left(\delta _{0}-R\,\delta _{0}^2\,\pi _{1}\right)}^2}{R\,\delta _{0}^2\,\pi _{1}^2-2\,\delta _{0}\,\pi _{1}+1}-\frac{m_{1}\,\left(4\,R\,\pi _{1}-2\right)}{\pi _{1}}\\&-&\frac{\delta _{0}^2\,n_{2}}{\delta _{0}\,\pi _{1}-1}+\frac{4\,m_{1}\,{\left(R\,\pi _{1}-1\right)}^2}{R\,\pi _{1}^2-2\,\pi _{1}+1}+4\,R\,\delta _{0}^2\,m_{2}-\frac{R\,m_{1}\,\left(4\,R\,\pi _{1}-2\right)}{R\,\pi _{1}-1}-\frac{2\,\delta _{0}\,m_{2}\,\left(2\,R\,\delta _{0}\,\pi _{1}-1\right)}{\pi _{1}}\\&-&\frac{2\,R\,\delta _{0}^2\,m_{2}\,\left(2\,R\,\delta _{0}\,\pi _{1}-1\right)}{R\,\delta _{0}\,\pi _{1}-1}, \\
I_{\pi_1,R}&=& E\left(-{\partial^2 l\over\partial \pi_1\partial R}\right) =     \frac{m_{1}\,\pi _{1}^2\,\left(2\,R\,\pi _{1}-2\right)}{R\,\pi _{1}^2-2\,\pi _{1}+1}-\frac{2\,R\,m_{1}\,\pi _{1}^2}{R\,\pi _{1}-1}+\frac{2\,\delta _{0}^3\,m_{2}\,\pi _{1}^2\,\left(R\,\delta _{0}\,\pi _{1}-1\right)}{R\,\delta _{0}^2\,\pi _{1}^2-2\,\delta _{0}\,\pi _{1}+1}-\frac{2\,R\,\delta _{0}^3\,m_{2}\,\pi _{1}^2}{R\,\delta _{0}\,\pi _{1}-1}, \\
  I_{R,R}&=&E\left(-{\partial^2 l\over\partial R^2}\right) =\frac{m_{1}\,\pi _{1}^2}{R}-\frac{2\,m_{1}\,\pi _{1}^3}{R\,\pi _{1}-1}+\frac{m_{1}\,\pi _{1}^4}{R\,\pi _{1}^2-2\,\pi _{1}+1}+\frac{\delta _{0}^2\,m_{2}\,\pi _{1}^2}{R}+\frac{\delta _{0}^4\,m_{2}\,\pi _{1}^4}{R\,\delta _{0}^2\,\pi _{1}^2-2\,\delta _{0}\,\pi _{1}+1}-\frac{2\,\delta _{0}^3\,m_{2}\,\pi _{1}^3}{R\,\delta _{0}\,\pi _{1}-1}.\\
\end{eqnarray*}

\subsection{Information matrix derivation for score CI and Wald-type CI}\label{appendix_derive_scoreCI_WaldCI}\hfill\\
The entries of the information matrix for $(\delta,\pi_1,R)$ has the form
$$I(\delta,\pi_1,R)= 
\begin{bmatrix}
  I_{11} & I_{12} & I_{13}\\
 I_{21} & I_{22} & I_{23} \\
 I_{31} & I_{32} & I_{33}\\
\end{bmatrix}
,$$
where
\begin{eqnarray*}  
I_{11}&=&E\left(-{\partial^2 l\over\partial \delta^2}\right)= \frac{n_{2}\,\pi _{1}}{\delta }+\frac{4\,m_{2}\,{\left(\pi _{1}-R\,\delta \,\pi _{1}^2\right)}^2}{R\,\delta ^2\,\pi _{1}^2-2\,\delta \,\pi _{1}+1}-\frac{n_{2}\,\pi _{1}^2}{\delta \,\pi _{1}-1}+4\,R\,m_{2}\,\pi _{1}^2-\frac{2\,m_{2}\,\pi _{1}\,\left(2\,R\,\delta \,\pi _{1}-1\right)}{\delta }\\&-&\frac{2\,R\,m_{2}\,\pi _{1}^2\,\left(2\,R\,\delta \,\pi _{1}-1\right)}{R\,\delta \,\pi _{1}-1},\\
I_{12}&=&E\left(-{\partial^2 l\over\partial \delta\partial\pi_1}\right)=n_{2}-m_{2}\,\left(4\,R\,\delta \,\pi _{1}-2\right)+m_{2}\,\left(8\,R\,\delta \,\pi _{1}-2\right)-\frac{m_{2}\,\left(2\,\pi _{1}\,\left(R\,\delta \,\pi _{1}-1\right)+2\,R\,\delta \,\pi _{1}^2\right)}{\pi _{1}}\\&+&\frac{m_{2}\,\left(2\,\delta -2\,R\,\delta ^2\,\pi _{1}\right)\,\left(2\,\pi _{1}-2\,R\,\delta \,\pi _{1}^2\right)}{R\,\delta ^2\,\pi _{1}^2-2\,\delta \,\pi _{1}+1}-\frac{\delta \,n_{2}\,\pi _{1}}{\delta \,\pi _{1}-1}-\frac{R\,\delta \,m_{2}\,\left(2\,\pi _{1}\,\left(R\,\delta \,\pi _{1}-1\right)+2\,R\,\delta \,\pi _{1}^2\right)}{R\,\delta \,\pi _{1}-1},\\
I_{13}&=&E\left(-{\partial^2 l\over\partial \delta\partial R}\right)=    -\frac{2\,\delta ^2\,m_{2}\,\pi _{1}^3\,\left(R-1\right)}{R^2\,\delta ^3\,\pi _{1}^3-3\,R\,\delta ^2\,\pi _{1}^2+R\,\delta \,\pi _{1}+2\,\delta \,\pi _{1}-1},\\
I_{21}&=&E\left(-{\partial^2 l\over\partial\pi_1 \partial\delta}\right)= n_{2}-m_{2}\,\left(4\,R\,\delta \,\pi _{1}-2\right)+m_{2}\,\left(8\,R\,\delta \,\pi _{1}-2\right)-\frac{m_{2}\,\left(2\,\delta \,\left(R\,\delta \,\pi _{1}-1\right)+2\,R\,\delta ^2\,\pi _{1}\right)}{\delta }\\&+&\frac{m_{2}\,\left(2\,\delta -2\,R\,\delta ^2\,\pi _{1}\right)\,\left(2\,\pi _{1}-2\,R\,\delta \,\pi _{1}^2\right)}{R\,\delta ^2\,\pi _{1}^2-2\,\delta \,\pi _{1}+1}-\frac{\delta \,n_{2}\,\pi _{1}}{\delta \,\pi _{1}-1}-\frac{R\,m_{2}\,\pi _{1}\,\left(2\,\delta \,\left(R\,\delta \,\pi _{1}-1\right)+2\,R\,\delta ^2\,\pi _{1}\right)}{R\,\delta \,\pi _{1}-1},\\
I_{22}&=&E\left(-{\partial^2 l\over\partial \pi_1^2}\right)=    4\,R\,m_{1}+\frac{n_{1}}{\pi _{1}}-\frac{n_{1}}{\pi _{1}-1}+\frac{\delta \,n_{2}}{\pi _{1}}+\frac{4\,m_{2}\,{\left(\delta -R\,\delta ^2\,\pi _{1}\right)}^2}{R\,\delta ^2\,{\pi _{1}}^2-2\,\delta \,\pi _{1}+1}-\frac{m_{1}\,\left(4\,R\,\pi _{1}-2\right)}{\pi _{1}}\\&-&\frac{\delta ^2\,n_{2}}{\delta \,\pi _{1}-1}+\frac{4\,m_{1}\,{\left(R\,\pi _{1}-1\right)}^2}{R\,{\pi _{1}}^2-2\,\pi _{1}+1}+4\,R\,\delta ^2\,m_{2}-\frac{R\,m_{1}\,\left(4\,R\,\pi _{1}-2\right)}{R\,\pi _{1}-1}-\frac{2\,\delta \,m_{2}\,\left(2\,R\,\delta \,\pi _{1}-1\right)}{\pi _{1}}\\&-&\frac{2\,R\,\delta ^2\,m_{2}\,\left(2\,R\,\delta \,\pi _{1}-1\right)}{R\,\delta \,\pi _{1}-1},\\
I_{23}&=&E\left(-{\partial^2 l\over\partial\pi_1 \partial R}\right)= 2\,m_{1}\,\pi _{1}+2\,\delta ^2\,m_{2}\,\pi _{1}+\frac{m_{1}\,\pi _{1}^2\,\left(2\,R\,\pi _{1}-2\right)}{R\,\pi _{1}^2-2\,\pi _{1}+1}-\frac{m_{1}\,\pi _{1}\,\left(4\,R\,\pi _{1}-2\right)}{R\,\pi _{1}-1}\\&-&\frac{2\,\delta ^2\,m_{2}\,\pi _{1}\,\left(2\,R\,\delta \,\pi _{1}-1\right)}{R\,\delta \,\pi _{1}-1}+\frac{2\,\delta ^3\,m_{2}\,\pi _{1}^2\,\left(R\,\delta \,\pi _{1}-1\right)}{R\,\delta ^2\,\pi _{1}^2-2\,\delta \,\pi _{1}+1},\\
I_{31}&=&E\left(-{\partial^2 l\over\partial R \partial\delta}\right)=    -\frac{\delta ^2\,m_{2}\,\pi _{1}^2\,\left(2\,\pi _{1}-2\,R\,\delta \,\pi _{1}^2\right)}{R\,\delta ^2\,\pi _{1}^2-2\,\delta \,\pi _{1}+1}-\frac{2\,R\,\delta ^2\,m_{2}\,\pi _{1}^3}{R\,\delta \,\pi _{1}-1},\\
I_{32}&=&E\left(-{\partial^2 l\over\partial R \partial\pi_1}\right)=    \frac{m_{1}\,\pi _{1}^2\,\left(2\,R\,\pi _{1}-2\right)}{R\,\pi _{1}^2-2\,\pi _{1}+1}-\frac{2\,R\,m_{1}\,\pi _{1}^2}{R\,\pi _{1}-1}+\frac{2\,\delta ^3\,m_{2}\,\pi _{1}^2\,\left(R\,\delta \,\pi _{1}-1\right)}{R\,\delta ^2\,\pi _{1}^2-2\,\delta \,\pi _{1}+1}-\frac{2\,R\,\delta ^3\,m_{2}\,\pi _{1}^2}{R\,\delta \,\pi _{1}-1},\\
I_{33}&=&E\left(-{\partial^2 l\over\partial R^2}\right)=    \frac{m_{1}\,\pi _{1}^2}{R}-\frac{2\,m_{1}\,\pi _{1}^3}{R\,\pi _{1}-1}+\frac{m_{1}\,\pi _{1}^4}{R\,\pi _{1}^2-2\,\pi _{1}+1}+\frac{\delta ^2\,m_{2}\,\pi _{1}^2}{R}+\frac{\delta ^4\,m_{2}\,\pi _{1}^4}{R\,\delta ^2\,\pi _{1}^2-2\,\delta \,\pi _{1}+1}-\frac{2\,\delta ^3\,m_{2}\,\pi _{1}^3}{R\,\delta \,\pi _{1}-1}.\\
\end{eqnarray*}  
Therefore, the upper left block of the inverse of the Fisher information matrix of $(\delta,\pi_1,R)$ is given by
$$
I^{\delta \delta}=\left(I_{11}-
    \begin{bmatrix}
      I_{12} & I_{13} 
    \end{bmatrix}
    \begin{bmatrix}
      I_{22} & I_{23} \\
      I_{32} & I_{33} 
    \end{bmatrix}^{-1}
    \begin{bmatrix}
      I_{21} \\
      I_{31}
    \end{bmatrix}
    \right)^{-1}.
$$
\end{document}